\documentclass[12pt]{oxarticle}
\usepackage{graphicx, float, array, xspace, amscd, amsmath, amsthm, amssymb, cite}

\let\SS=\S 
\pdfoutput=1 
\newcommand{\varstr}[2]{\vrule height #1 depth #2 width0pt}

\newcommand{\D}{\Delta}

\renewcommand{\S}{\Sigma}

\newcommand{\ch}{\chi}

\newcommand{\cM}{\mathcal{M}}

\newcommand{\cV}{\mathcal{V}}

\newcommand{\IP}{\mathbb{P}}

\newcommand{\IR}{\mathbb{R}}

\newcommand{\IZ}{\mathbb{Z}}

\font\eightrm=cmr8 at 8pt
\font\csc=cmcsc10

\newcommand{\beq}{\begin{equation}}
\newcommand{\eeq}{\end{equation}}
\newcommand{\beqnn}{\begin{equation*}}
\newcommand{\eeqnn}{\end{equation*}}

\newcommand{\fref}[1]{Fig.~\!\ref{#1}}
\newcommand{\tref}[1]{Table~\ref{#1}}
\newcommand{\sref}[1]{\SS\ref{#1}}     

\newcommand{\hodgenos}{(h^{1,1},\,h^{2,1})}

\def\place#1#2#3{\vbox to0pt{\kern-\parskip\kern-7pt
                             \kern-#2truein\hbox{\kern#1truein #3}
                             \vss}\nointerlineskip}
                        
\newcommand{\capt}[3]{\parbox{#1}{\renewcommand{\baselinestretch}{1.0}
                                                           \caption{\label{#2}\small\it #3}}}

\newcommand{\cys}{Calabi-Yau manifolds\xspace}

\newcommand{\SDelta}{\Delta^{^{\!\!*}}}

\newcommand{\Top}[1]{\langle #1 |}    
\newcommand{\Bot}[1]{| #1 \rangle}    
\newcommand{\topbot}[2]{\langle #1 | #2 \rangle}

\renewcommand{\baselinestretch}{1.1}
\numberwithin{equation}{section}
\proofmodefalse

\begin{document}
\pagestyle{empty}
\begin{center}
\null\vskip0.2in
{\Huge An Abundance of K3 Fibrations\\ [0.1in]   
from \\ [0.1in]
Polyhedra with Interchangeable Parts\\[0.4in]}
{\csc Philip Candelas$^{\,1}$, Andrei Constantin$^{\,2}$ \\
and\\
Harald Skarke$^{\,3}$\\[0.3in]}
{\it $^1$Mathematical Institute\hphantom{$^1$}\\
University of Oxford\\
24-29 St.\ Giles'\\
Oxford OX1 3LB, UK\\[0.2in]
$^2$Rudolf Peierls Centre for Theoretical Physics\hphantom{$^2$}\\
University of Oxford\\
1 Keble Road\\
Oxford OX1 4NP, UK\\[0.2in]
$^3$Institute for Theoretical Physics\hphantom{$^2$}\\
Vienna University of Technology\\
Wiedner Hauptstrasse 8-10/136\\
1040 Vienna, Austria\\}
\footnotetext[1]{candelas@maths.ox.ac.uk \hfill 
$^2\,$a.constantin1@physics.ox.ac.uk \hfill
$^3\,$skarke@hep.itp.tuwien.ac.at}
\vfill
{\bf Abstract\\}
\end{center}
\vspace*{-15pt}
Even a cursory inspection of the Hodge plot associated with Calabi-Yau threefolds that are hypersurfaces in toric varieties reveals striking structures. These patterns correspond to webs of elliptic-$K3$ fibrations whose mirror images are also elliptic-$K3$ fibrations. Such manifolds arise from reflexive polytopes that can be cut into
two parts along slices corresponding to the $K3$ fibers. Any two half-polytopes over a given slice can be combined into a reflexive polytope. This fact, together with a remarkable relation on the additivity of Hodge numbers, explains much of the structure of the observed~patterns.
%
%
\newpage
\tableofcontents
\newpage
\setcounter{page}{1}
\pagestyle{plain}
\rightline{\it This paper is dedicated to the memory of Max Kreuzer.}
\vskip20pt
\section{Introduction}
\vskip-10pt
To date, the largest class of Calabi-Yau threefolds that has been constructed explicitly, consists of hypersurfaces in toric varieties which are associated to reflexive polytopes via the Batyrev construction~\cite{Batyrev:1993dm}. Kreuzer and the third author have given a complete list of 473,800,776 such polytopes~\cite{Kreuzer:2000xy, Kreuzer:2000qv}. The Hodge numbers $h^{1,1}$ and $h^{1,2}$ play an important role in the classification of Calabi-Yau manifolds and in applications of these manifolds to string theory. 
There are combinatorial formulas for these numbers in terms of the polytopes, that are given in~\cite{Batyrev:1993dm}. By computing the Hodge numbers associated to the polytopes in the list, one obtains a list of 30,108 distinct pairs of values for $\hodgenos$. 
\begin{figure}[H]
\begin{center}
\includegraphics[width=6.5in]{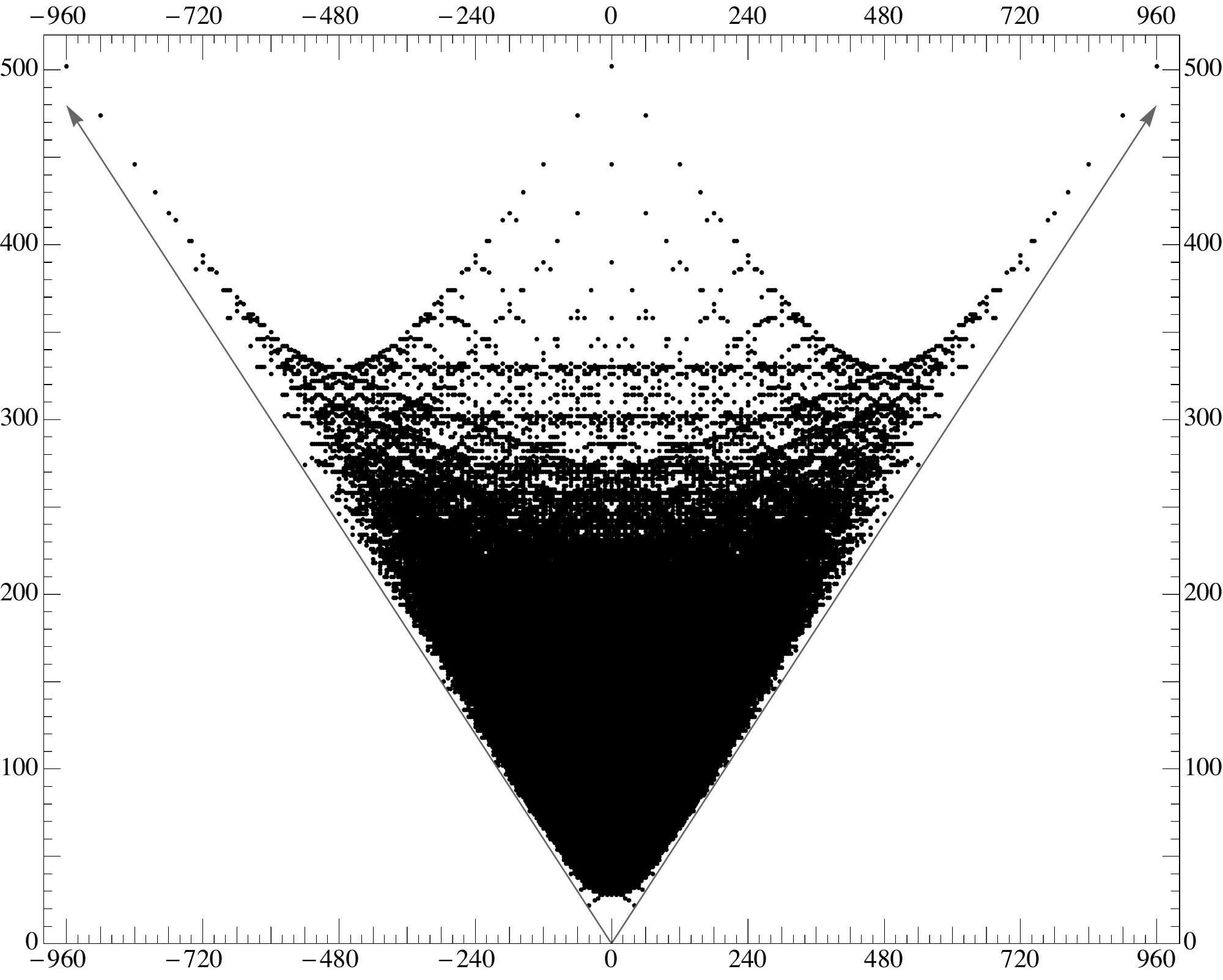}
\capt{6.5in}{BasicKSPlot}{The Hodge plot for the list or reflexive 4-polytopes. The Euler number
$\chi = 2\left( h^{1,1}-h^{1,2} \right)$ is plotted against the height $y=h^{1,1}+h^{1,2}$. The oblique axes correspond to $h^{1,1}=0$ and $h^{1,2}=0$.}
\end{center}
\end{figure}
These are presented in \fref{BasicKSPlot}, in which we plot the Euler number 
$\chi = 2\left( h^{1,1}-h^{1,2} \right)$ against what we shall call the height, $y=h^{1,1}+h^{1,2}$.
The plot has an intriguing structure. One immediate feature of this plot, also evident from Batyrev's formulae, is the presence of mirror symmetry at the level of Hodge numbers: the Hodge numbers associated to a reflexive polytope are interchanged with respect to the dual polytope. This corresponds to the symmetry about the axis 
$\chi=0$. 
\begin{figure}[!t]
\begin{center}
\includegraphics[width=6.5in]{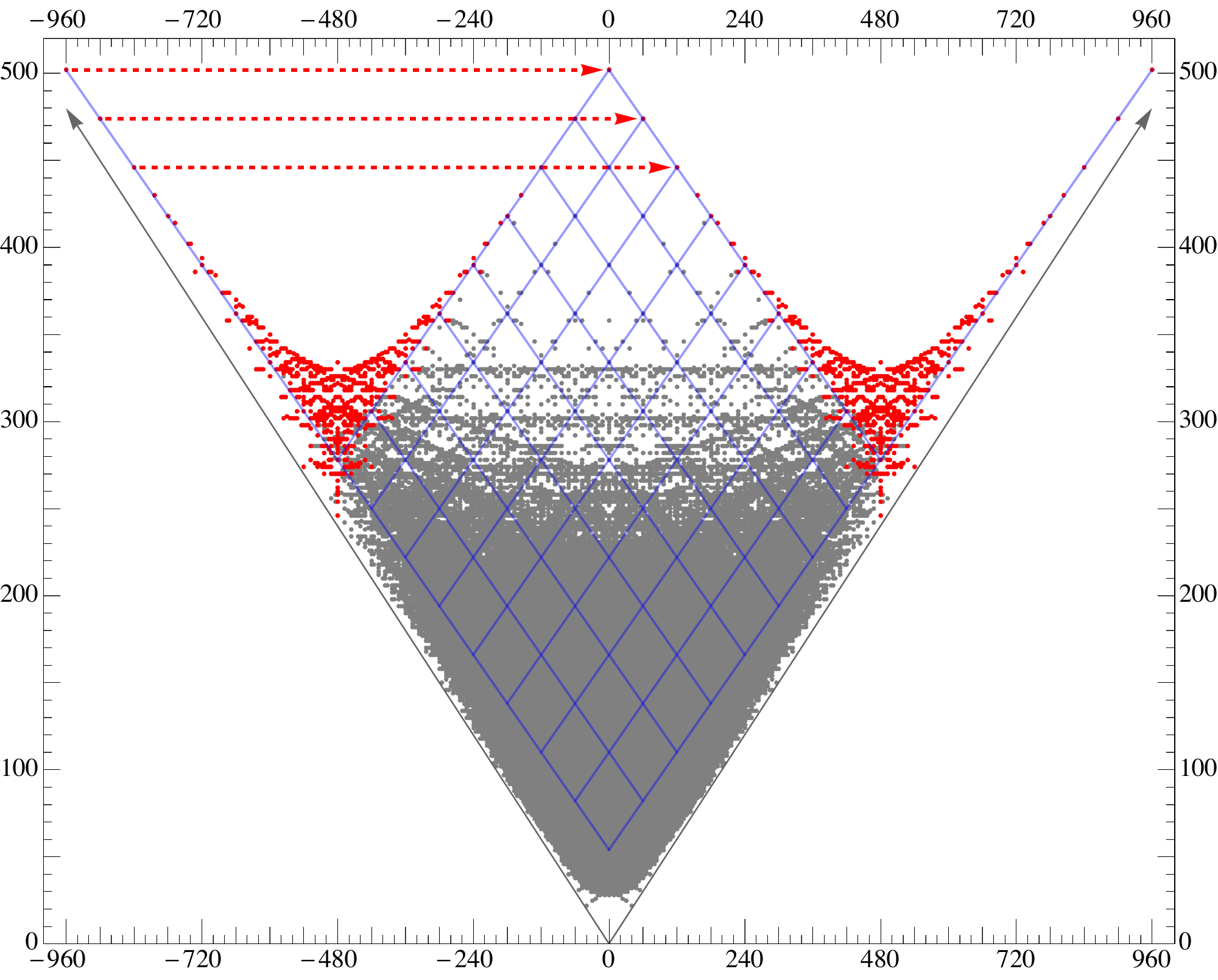}
\capt{6.2in}{YStructure}{The structure in red, on the left, contains points which have mirror images when reflected about the axis $\chi = -480$. The red arrows show that these points can be mapped into points corresponding to manifolds with positive Euler number by a change in Hodge numbers $\Delta\big(h^{1,1},h^{1,2} \big)=(240,-240)$, corresponding to $\D(\chi, y)=(960,0)$.}
\end{center}
\vskip-10pt
\end{figure}
Another striking feature is that both the left and the right hand sides of the plot contain structures symmetric about vertical lines corresponding to Euler numbers $\chi= \mp 480$. One can easily observe that the great majority of the points corresponding to manifolds with $\chi<-480$ have mirror images when reflected about the $\chi=-480$ axis. In \fref{YStructure},  
the structure which exhibits this half-mirror symmetry is highlighted in red. Equivalently, one can observe that the red points, on the left, and only those, can be translated into other points of the plot, corresponding to manifolds with positive Euler number, by a change in Hodge numbers 
$\D(h^{1,1},h^{1,2} )=(240,-240)$ corresponding to 
$\D(\chi, y)=(960,0)$, as indicated by the red arrows in \fref{YStructure}. Together with mirror symmetry, this results in the symmetry described above.

\begin{figure}[!bt]
\begin{center}
\includegraphics[width=6.5in]{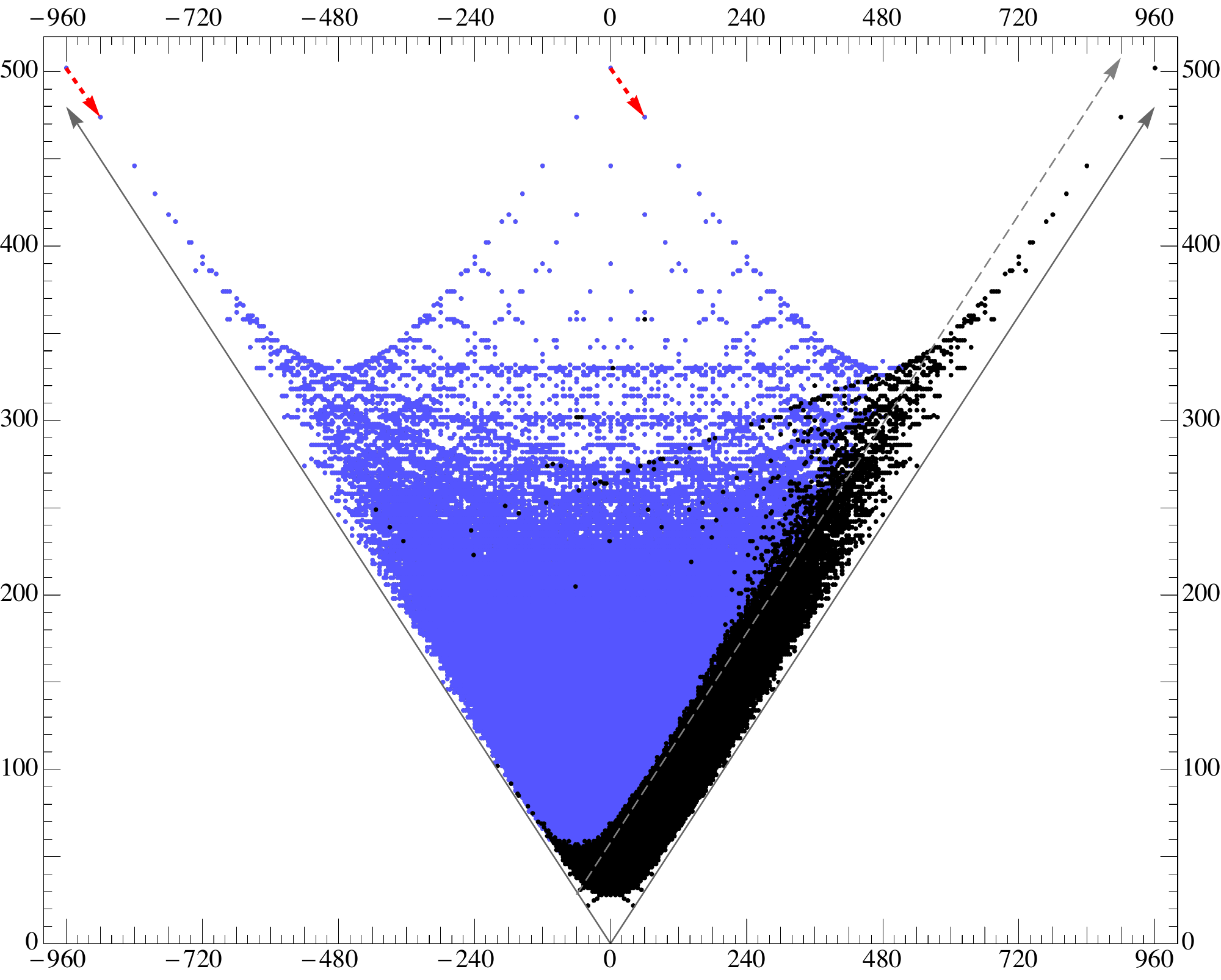}
\capt{5.8in}{BlueVectorPlot}{The blue points can be translated into other points of the plot by a change in Hodge numbers $\Delta\big(h^{1,1},h^{1,2} \big)=(1,-29)$. The red arrows illustrate this action on two pairs of points. The extra grey arrow corresponds to $h^{1,2} = 30$.}
\end{center}
\vskip-10pt
\end{figure}

Yet another intriguing feature is evident from \fref{YStructure}: a special role is played by  a vector $\Delta\big(h^{1,1},h^{1,2} \big)=(1,-29)$, corresponding to $\D(\chi,y)=(60,-28)$. This, together with its mirror, are the displacements corresponding to the blue grid. It is immediately evident that many points have a `right descendant' corresponding to these displacements. However, it is a fact that almost  	all points with $h^{1,2} \geq 30$ have such descendants. In \fref{BlueVectorPlot} we have coloured each point, that has a right-descendant, blue.  Note also how these translations, together with their mirrors, account for the gridlike structure in the vicinity of the central peak of~the~plot.
\goodbreak
 
One can consider also `left-descendants', that is points of the plot that are displaced by the mirror vector, 
$\D(\chi,y)=(-60, -28)$, from a given point. There are very few points of the plot that do not have either a left or right descendant, as illustrated by \fref{NoDescendants}.

\begin{figure}[!t]
\begin{center}
\includegraphics[width=6.5in]{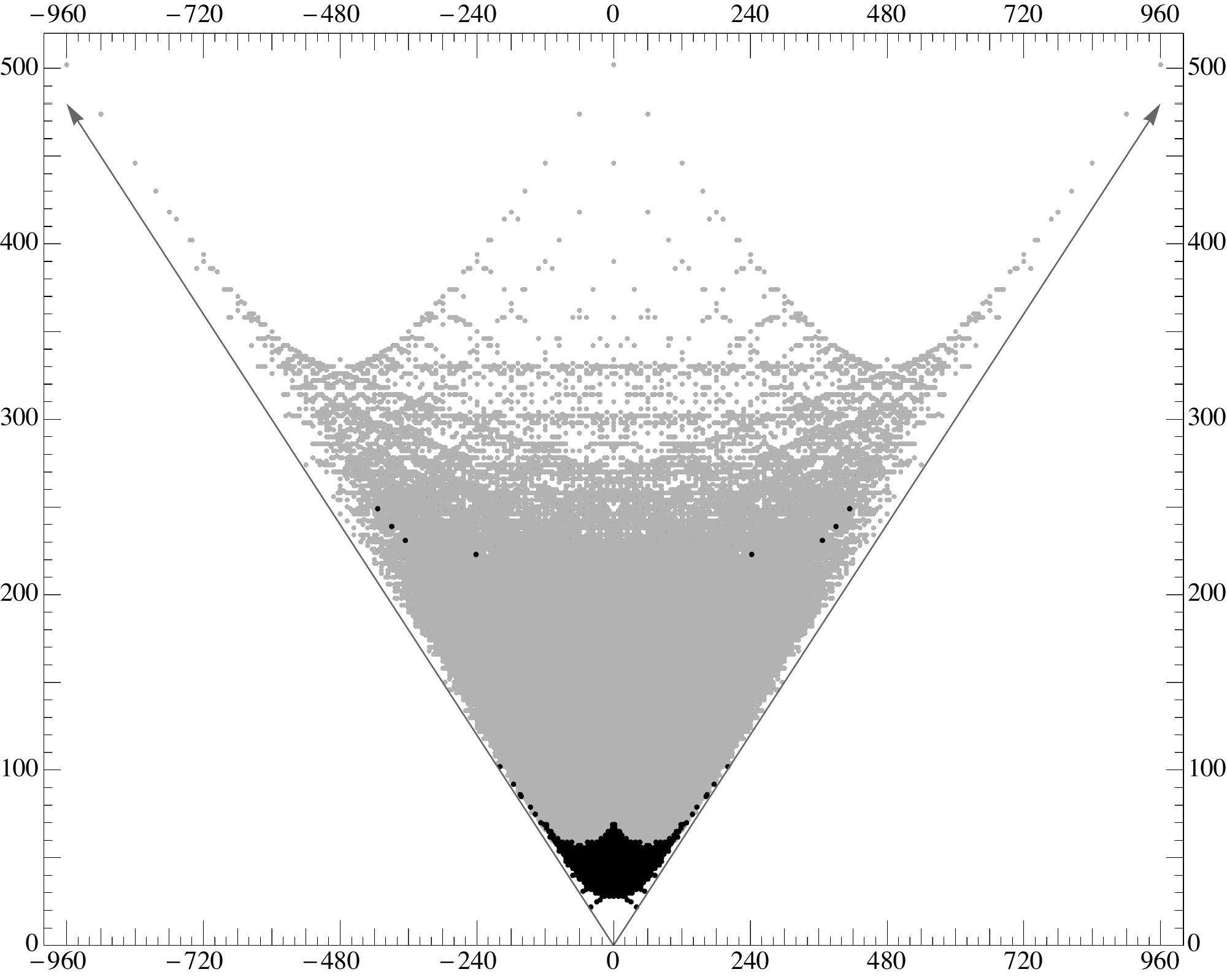}
\capt{6.3in}{NoDescendants}{The points in black are the only ones that have neither a left nor a right descendant.}
\end{center}
\vskip-14pt
\end{figure}

The aim of this paper is to study the structures discussed above. Our starting point is the observation that most of the points making up the red structure in \fref{YStructure} 
correspond to Calabi-Yau threefolds fibered by $K3$ surfaces, which are themselves elliptically fibered. This nested fibration structure is visible in the reflexive polytopes which provide the toric description of these Calabi-Yau manifolds. 
As we will recall in \sref{NestedK3Fibrations}, the 4-dimensional reflexive polytope $\SDelta$ associated to a toric Calabi-Yau threefold describes a $K3$ fibration if there exists a 3-dimensional reflexive polyhedron (associated with the $K3$ fiber) that is contained in the 4-dimensional polytope as a slice~\cite{Avram:1996pj}. Hence the $K3$ polyhedron naturally divides the 4-dimensional polytope into two parts, a `top' and a `bottom'. For a given $K3$ polyhedron there exists a finite number of different tops and bottoms that can be assembled into a reflexive 4-dimensional polytope. 
\begin{figure}[!t]
\begin{center}
\includegraphics[width=6.5in]{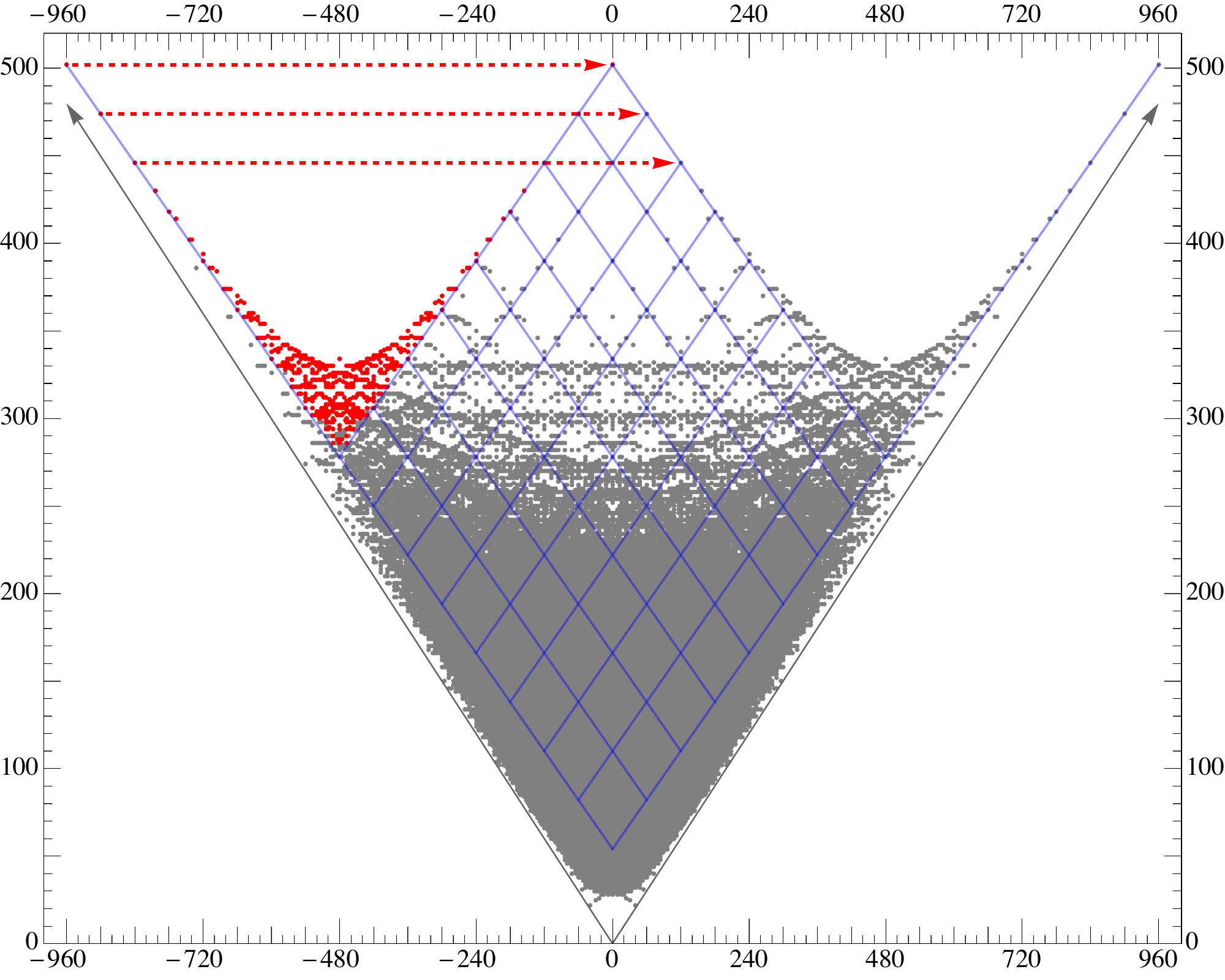}
\capt{6.5in}{BigBotPoints}{The 465 pairs of Hodge numbers that result from combining the 1263 possible $E_8{\times}\{1\}$ tops with the maximal bottom. Note that this structure is contained in, though is not identical to, the structure shown on the left in \fref{YStructure}. Exchanging the maximal bottom for the minimal bottom shifts the entire structure by the vector $\D(\chi,y)=(960,0)$, corresponding to the red arrows.}
\end{center}
\vskip-10pt
\end{figure}

\subsection{The role of the $E_8{\times}\{1\}$ K3 polyhedron}
Consider the mirror pair $\big(\cM_{11,491}$, $\cM_{491,11}\big)$ of Calabi-Yau manifolds occupying the top left and top right positions in the Hodge plot. The polytope $\Delta_{11,491}=\D^{^{\!\!*}}_{491,11}$ is actually the largest (that is, it has the largest number, 680, of lattice points) of all reflexive 4-polytopes and admits two distinct slicings along $K3$ polytopes. In the first case the $K3$ polytope is the largest reflexive 3-polytope (with 39 lattice points), which can be associated with a gauge group of either $E_8{\times} E_8$~\cite{Candelas:1996su} or $SO(32)$~\cite{Candelas:1997pq}; slicing along this $K3$ polytopes leads to the largest known gauge groups coming from F-theory compactification~\cite{Candelas:1997eh,Candelas:1997pq}.
In the present paper, however, we are concerned with the second case, where the $K3$ polytope is self-dual and corresponds to $E_8{\times}\{ 1\}$ (for the classification of elliptic-$K3$ polyhedra by Lie groups see, for example,~\cite{Candelas:1996su, Bouchard:2003bu}).
Moreover, the top and the bottom are the same and they are the largest among all the available tops and bottoms for that $K3$ manifold.  The mirror is also an elliptic $K3$ fibration. Its polytope is divided into a top and a bottom by the same slice, though now this top (which is again the same as the bottom) is the smallest  among all the available tops and~bottoms. 
By taking arbitrary tops, from the ones which fit this $K3$ slice, together with the biggest bottom we are able to obtain many elliptically fibered \cys . These give rise to the red points that form the $V$-shaped structure on the left of \fref{BigBotPoints}. Exchanging the maximal bottom with the smallest one, while keeping the top fixed, shifts the Hodge numbers by $\Delta\big(h^{1,1},h^{1,2} \big)=(240,-240)$,  in terms of the coordinates of the plot this is $\D(\chi,y)=(960,0)$, corresponding to the red arrow of~the~figure.

There are 1,263 different tops, and so also 1,263 bottoms that project onto this $K3$ slice,
and these can be attached along the $K3$ polytope to obtain $1263{{\times}}1264/2 = 798,216$ \cys which are elliptic-$K3$ fibrations. This already gives us the largest collection of elliptic-$K3$ fibrations known hitherto. There are many different $K3$ polyhedra to which this construction can be applied, only a few of which we will discuss here.

The Hodge numbers of the manifolds obtained by combinations of the 1263 tops and bottoms discussed above are related by a remarkable formula. Namely, if we denote by $\Top{A}$ and $\Top{C}$ two of these tops and by 
$\Bot{B}$ and $\Bot{D}$ two of the bottoms, and we write $\topbot{A}{B}$ for the polytope formed by joining $\Top{A}$ and $\Bot{B}$ along their common base, then the Hodge numbers $h^{1,1}$ and $h^{1,2}$ satisfy the relation:
\beq
h^{\bullet\bullet}\big(\topbot{A}{B}\big)+h^{\bullet\bullet}\big(\topbot{C}{D}\big)~=~
h^{\bullet\bullet}\big(\topbot{A}{D}\big)+h^{\bullet\bullet}\big(\topbot{C}{B}\big)~.
\label{hnosrelation}\eeq
The 1,263 tops, when combined with the maximal bottom give rise to a set of 465 pairs of Hodge numbers. These are the points  shown in red in \fref{BigBotPoints}. We shall refer to this set of points as the $V$-structure. The relation \eqref{hnosrelation} has an important consequence for these points. Consider any vector taking the pair (11,491) to one of the remaining 464 points. The relation above ensures that we may translate the entire \hbox{$V$-structure} by each of these 464 vectors. Translating the $V$-structure by these vectors explains much of the repetitive structure associated with the blue grid~of~\fref{BigBotPoints}. It also enables us to calculate the Hodge numbers of the resulting Calabi-Yau manifolds. In this way, we find 16,148 distinct pairs of Hodge numbers. The result of performing all 464 translations on the $V$-structure is~shown~in~\fref{AllE8TimesSU1}.

The vectors with which our considerations began are included in the $V$-structure translations. The vector $\D\hodgenos=(240,-240)$ arises as the difference between $\hodgenos=(11,491)$ and 
$\hodgenos=(251,251)$. We know that $\D\hodgenos=(1,-29)$ appears among the translations, in fact the vectors $\D\hodgenos=k{\times}(1,-29)$ appear  for $1\leq k\leq 7$. 
Finally, the blue grid~of~\fref{BigBotPoints} closes due to the fact that the horizontal shift $\Delta\big(h^{1,1},h^{1,2} \big)=(240,-240)$ is related to the 
$\D(h^{11},h^{21})=(1,-29)$ shift and its mirror reflection by 
$$
(240,-240) ~=~ 8{\times}(1,-29) - 8{\times}(-29,1)~.
$$
Note, however, that the vector $8{\times}(1,-29)$ is not, itself, a $V$-structure translation. 
\begin{figure}[!t]
\begin{center}
\includegraphics[width=6.5in]{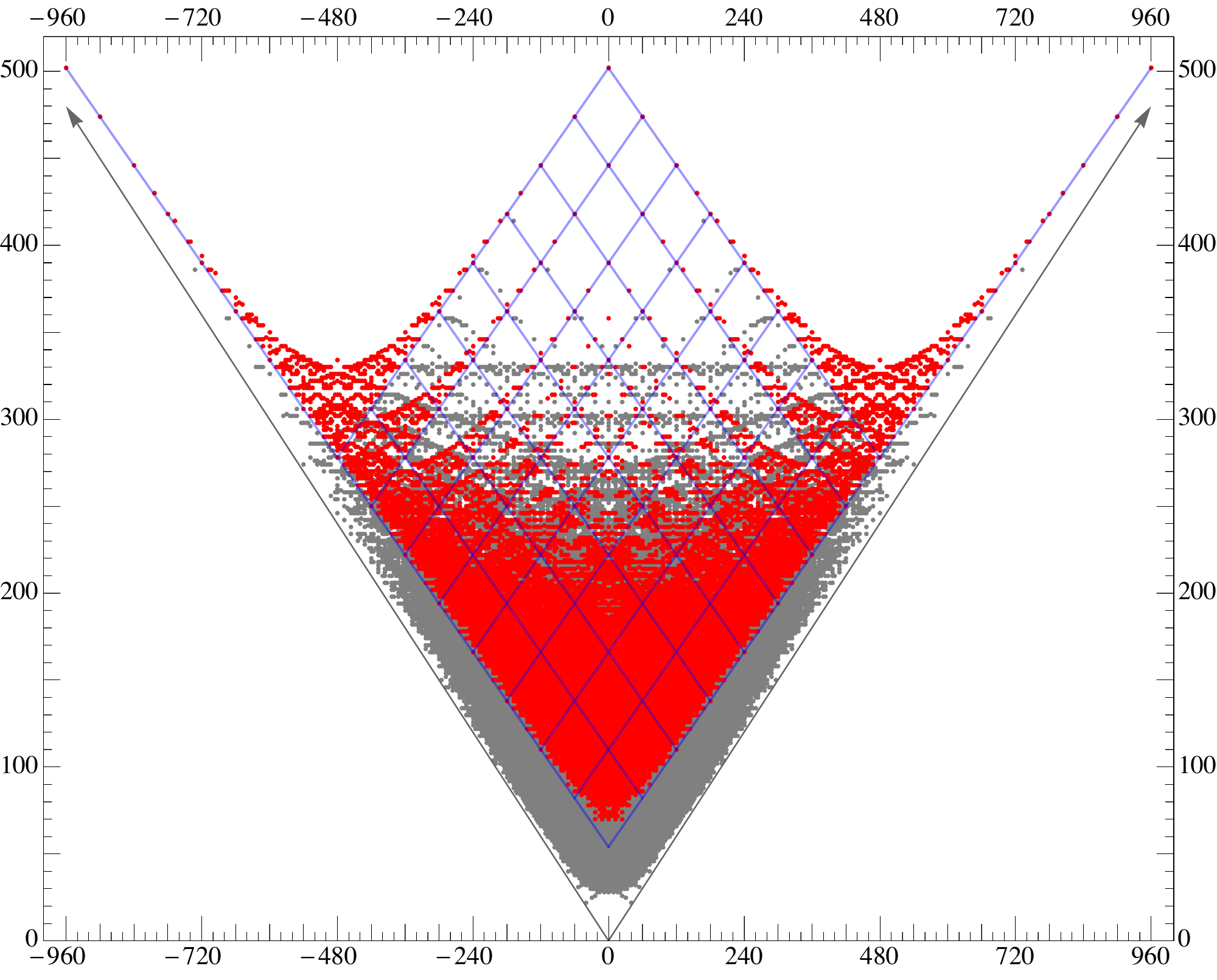}
\capt{5.5in}{AllE8TimesSU1}{The Hodge plot showing, in red, all $E_8{\times}\{1\}$ $K3$ fibrations for which the fiber is both a slice and a projection in the four-dimensional reflexive polytope.}
\end{center}
\end{figure}

\subsection{Other K3 polyhedra}
There are many types of elliptically fibered $K3$ polyhedra.
These are associated with groups $G_1{\times}G_2$ which describe the way in which the elliptic fiber degenerates along the base, according to the ADE classification of singularities\footnote{The tops we use are associated with Lie groups. These groups need not be simply laced, indeed the groups $G_2$ and $F_4$ arise in our discussion. In the context of $F$-theory these groups arise, via monodromy, from the simply laced groups.}. 
Here the fibration structure manifests itself as a slice along a reflexive polygon, with corresponding three-dimensional tops and bottoms to which the gauge groups can be associated, as first observed 
in~\cite{Candelas:1996su}.
\begin{figure}[!t]
\begin{center}
\includegraphics[width=6.5in]{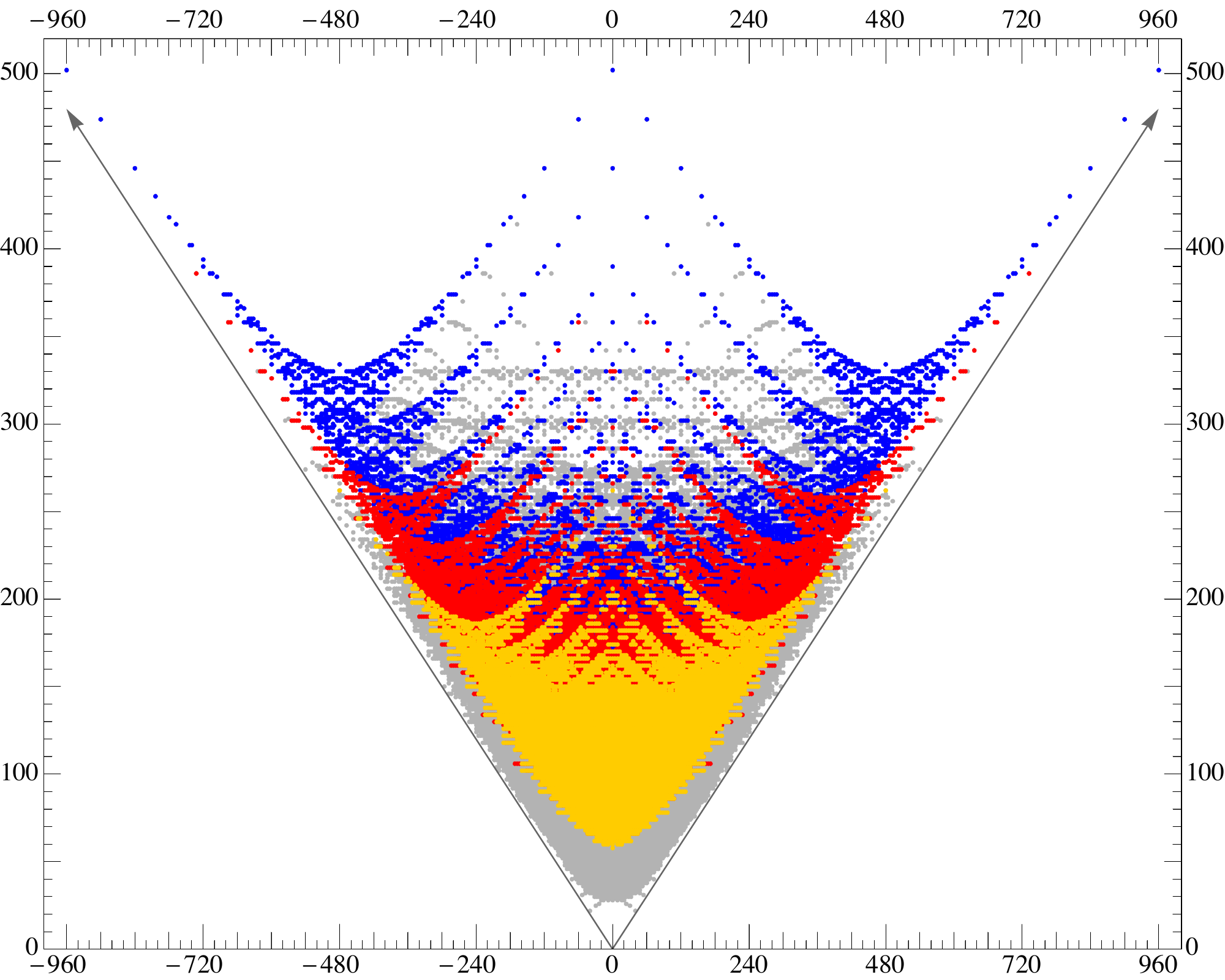}
\capt{5.3in}{RBGplot}{Hodge plot showing in blue all $E_8{\times} \{1\}$ $K3$ fibrations; in red all $E_7{\times} \{1\}$ and $E_8{\times} SU(2)$ $K3$ fibrations; and in yellow: all $E_7{\times} SU(2)$ $K3$ fibrations.}
\end{center}
\end{figure}
The complete lists of tops for any of the 16 reflexive polygons can be found in~\cite{Bouchard:2003bu}.
In the present paper we are only concerned with the case where the reflexive
polygon is the triangle corresponding to a Weierstrass model. Usually we also assume that both the $K3$ polyhedron and its dual are
subpolytopes of the polytope associated with $E_8\times E_8$. 
In this case the dual 3-polytope gives rise to $\widetilde G_1{\times}\widetilde G_2$,
where $\widetilde G_1$ is the commutant of $G_1$ in $E_8$, and likewise for $G_2$.
In particular, if $G_2$ is the commutant of $G_1$ in $E_8$, the $K3$ polyhedron is self-dual. The class of elliptic-$K3$ fibered Calabi-Yau manifolds discussed above belongs to this category and 
corresponds~to~$(G_1,G_2)=(E_8,\{1\})$. 

We have also studied the case $(G_1,G_2)=(E_7, SU(2))$ which we find very similar to the previous one and which gives rise to $725,410$ polyhedra and $7,929$ pairs of Hodge numbers. The $E_7{\times}SU(2)$ structure corresponds to the yellow points in \fref{RBGplot}. While similar to the $E_8\times\{1\}$ case, this structure is not immediately apparent in the original Hodge plot of \fref{BasicKSPlot} owing to the fact it appears in the region of the plot that is very dense.  

\goodbreak
We also generalise our construction to slices corresponding to non-self-dual $K3$ polyhedra by studying the case $(G_1,G_2)=(E_7, \{1\})$ and the corresponding dual $K3$ polyhedron with $(G_1,G_2)=(SU_2, E_8)$. The $E_7\times \{1\}$ structure and its mirror are presented in \fref{E7E8str}. The top part of these structures, which is not overlaid by the $E_7\times SU(2)$ structure corresponds to the red points in~\fref{RBGplot}. The $E_7\times \{1\}$ structure and its mirror are particularly interesting as they pick up some of the points of the structure presented in \fref{YStructure} that lie to the left of the $V$-structure.

\subsection{Layout of the paper}
The remainder of the paper is largely dedicated to the substantiation of the discussion of this introduction.
In \sref{NestedK3Fibrations} we provide the background necessary for analysing the webs of
nested fibration structures. We give the required definitions, discuss the connection of fibration structures with string dualities, and prove two lemmata concerning the compatibility of tops with bottoms and the additivity (\ref{hnosrelation}) of the corresponding Hodge numbers.
In \sref{E81} we discuss the web of $E_8{\times} \{1\}$ elliptic $K3$ fibrations, and in \sref{MoreWebs} we generalize the construction to the cases of $E_7{\times} SU(2)$ and $E_7{\times} \{1\}$.
Finally in \sref{Outlook} we summarise our results and discuss how one might proceed.
In particular, we identify several starting points for further webs.

\subsection{Lists of tops\label{webaddress}}
In the course of this work we have compiled lists of all tops for $E_8{\times}\{1\}$, $E_7{\times}SU(2)$ and $E_8{\times}SU(2)$ fibrations. The tops for $E_7{\times}\{1\}$ fibrations can be recovered from those for 
$E_8{\times}SU(2)$ by computing dual polytopes. For reasons of length, we do not give these lists here but have posted them to {\tt http://hep.itp.tuwien.ac.at/\raisebox{-7pt}{\Large\textasciitilde}skarke/NestedFibrations/}~.
\newpage
\section{Nested $K3$ Fibrations}\label{NestedK3Fibrations}
In this section we define the nested structures of reflexive polytopes which correspond to elliptic-$K3$ fibration structures of Calabi-Yau manifolds. We also briefly discuss the place of these fibration structures in F-theory/type IIA duality. We give a precise formulation and a proof of the additivity property, mentioned in the introduction, under the assumption that the $K3$ slice obeys a specific condition, which is satisfied by the principal examples that we discuss. We also give a counter-example to show that the relation need not apply otherwise.

\subsection{Reflexive polytopes and toric Calabi-Yau hypersurfaces} Let $N,M\simeq \mathbb Z^n$ be two dual lattices of rank $n$ and let 
$\langle \cdot ,\cdot \rangle:M\times N\rightarrow\mathbb Z$ 
denote the natural pairing. Define the real extensions of $N$ and $M$ as $N_{\mathbb R}:=N\otimes\, \mathbb R$ and $M_{\mathbb R}:=M\otimes\, \mathbb R$. A~polytope $\Delta \subset M_{\mathbb R}$ is defined as the convex hull of finitely many points in $M_{\mathbb R}$ (its vertices). The set of vertices of $\Delta$ is denoted by $\cV(\Delta)$ and its relative interior by $\text{int}(\Delta)$. A lattice polytope is a polytope for which $\cV(\Delta)\subset M$. Reflexivity of polytopes is a property defined for polytopes for which 
$\text{int}(\Delta)\cap M$ 
contains the origin. Then a lattice polytope is said to be reflexive if all its facets are at lattice distance 1 from the origin, that is there is no lattice plane, parallel to the given facet, that lies between the facet and the origin.

The polar, or dual polytope of a reflexive polytope is defined as the convex hull of inner normals to facets of~$\Delta$, normalised to primitive lattice points of $N$. Equivalently, a reflexive polytope $\Delta\subset M_{\mathbb R}$ is defined as a lattice polytope having the origin as a unique interior point, whose dual 
\begin{equation}
\SDelta = \{ y\in N_{\mathbb R}: \langle x,y\rangle \geq -1, \text{ for all } x \in \Delta \}
\end{equation}\label{dual}%
is also a lattice polytope. 

To any face $\theta$ of $\Delta$ one can assign a dual face $\theta^*$ of
$\SDelta$ as 
$$\theta^*=
\{ y\in\SDelta:\! \langle x,y\rangle = -1, \text{ for all } x \in \theta \}.$$
In this way a vertex is dual to a facet, an edge to a codimension 2 face and so on.
In particular, for 3-polytopes edges are dual to edges. 

One can construct a toric variety from the fan over a triangulation of the surface of $\SDelta$, and a Calabi-Yau hypersurface in this toric variety as the zero locus of a polynomial whose monomials are in one-to-one correspondence with the lattice points of $\Delta$. This construction is described in the texts~\cite{0813.14039, 1223.14001}. For an account in the spirit of the present paper see~\cite{Skarke:1998yk, Avram:1996pj}.

\subsection{Reflexive polytopes and $K3$-fibered Calabi-Yau manifolds}
There has been a long standing interest in $K3$-fibered Calabi-Yau threefolds in string theory. $K3$ fibrations appear in a natural way in the study of four dimensional $\mathcal N=2$ heterotic/type IIA duality~\cite{Klemm:1995tj, Vafa:1995gm, Aspinwall:1995vk}. Toroidal compactifications of the strongly coupled heterotic string theory to six dimensions are dual to weakly coupled type IIA theory compactifications on $K3$ surfaces, in the sense that the moduli spaces of vacua for both sides match. In~\cite{Vafa:1995gm} it was noted that this duality can be carried over to four dimensions if the the two theories are fibered over $\mathbb P^1$, that is if the type IIA theory is compactified on a manifold which is a $K3$ fibration over $\mathbb P^1$ and the heterotic string is compactfied on $K3{\times} T^2$, which can be written as a $T^4$ fibration over $\mathbb P^1$. In this way, the six dimensional heterotic/type IIA duality can be used fiber-wise. 

In the same context, Aspinwall and Louis~\cite{Aspinwall:1995vk} showed that, after requiring that the pre-potentials in the two theories match, and assuming that the type IIA theory is compactified on a Calabi-Yau manifold, this manifold must admit a $K3$ fibration. At the time, several lists of $K3$ fibered Calabi-Yau threefolds have been compiled, first as hypersurfaces in weighted projective 4-spaces~\cite{Klemm:1995tj, Hosono:1996ua} and shortly after that by using the methods of toric geometry~\cite{Avram:1996pj}. The language of toric geometry was also used in~\cite{Candelas:1996su} where the authors noticed that the Dynkin diagrams of the gauge groups appearing in the type IIA theory can be read off from the polyhedron corresponding to the $K3$ fibered Calabi-Yau manifold used in the compactification. The singularity type of the fiber corresponds to the gauge group in the low-energy type IIA theory. In the case of an elliptically fibered $K3$, the Dynkin diagrams appear in a natural way~\cite{Perevalov:1997vw}.

The discussion below uses the same language of reflexive polytopes. The fibration structure of a Calabi-Yau threefold is described by a pair $(\Delta, \SDelta) \subset M{\times}N$ of reflexive polytopes where $N$ has a distinguished three dimensional sub-lattice $N_3$, such that $\SDelta_3= \SDelta \cap N_3$ is a three dimensional reflexive polytope. The sub-polytope $\SDelta_3$ corresponds to the fiber and divides the polytope $\SDelta$ into two parts, a top and a bottom. The fan corresponding to the base space can be obtained by projecting the fan of the fibration along the linear space spanned by sub-polytope describing the fiber~\cite{Kreuzer:1997zg}. In the present case ($\SDelta$ a four-dimensional polytope and $\SDelta_3$ three dimensional), the base space is always $\mathbb P^1$. This description is dual to having a distinguished one dimensional sub-lattice $M_1 \subset M$, such that the projection of $\Delta$ along $M_1$ is $\Delta_3= (\SDelta_3)^{\!^*}$. The equivalence between the two descriptions has been proved in~\cite{Avram:1996pj}, and was expressed as `a slice is dual to a projection'. In the case where the mirror Calabi-Yau is a fibration over the mirror $K3$, it is possible to introduce distinguished three and, respectively, one dimensional sub-lattices $M_3$ and $N_1$, resulting in the splits $M = M_1 \oplus M_3$ and $N = N_1 \oplus N_3$.

\subsection{Nested elliptic-$K3$ Calabi-Yau fibrations}
This special type of fibration structure corresponds to Calabi-Yau manifolds which are $K3$ fibrations over $\mathbb P^1$ and for which the fiber is itself an elliptic fibration. Such manifolds appeared in~\cite{Candelas:1996su} in the discussion of heterotic/type IIA (F-theory) duality. In the toric language, such fibration structures are displayed in the form of nestings of the corresponding~polytopes. 

A degeneration of the elliptic fibration may lead to a singularity of ADE type which can be resolved by introducing a collection of exceptional divisors whose intersection pattern is determined by the corresponding A, D or E type Dynkin diagram.
As the divisors correspond to lattice points of $\SDelta$, the group
can be read off from the distribution of lattice points in the top and the bottom (above and below the polygon corresponding to the elliptic curve)~\cite{Candelas:1996su, Perevalov:1997vw}. 

\begin{figure}[!t]
\begin{center}
\framebox[6.5in][c]{\includegraphics[width=5.7in]{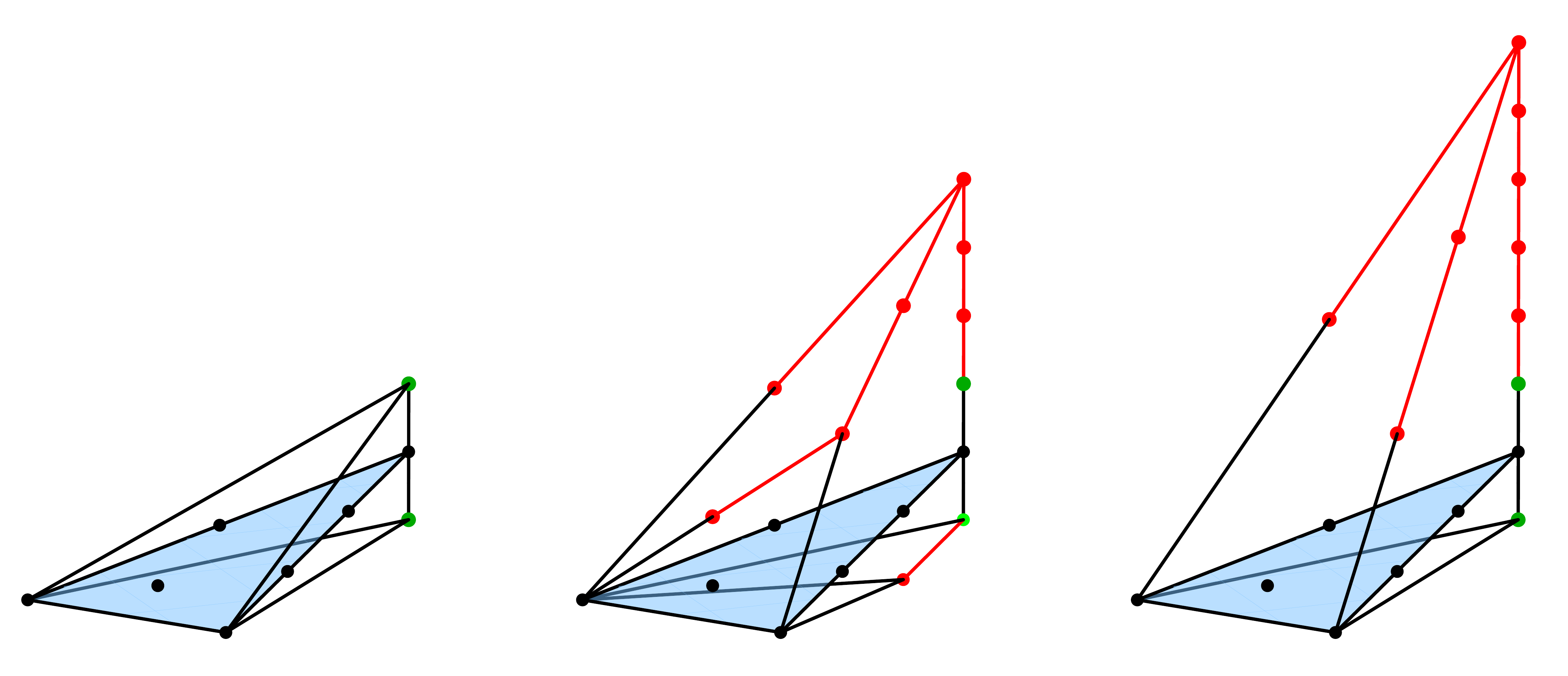}}
\capt{5.3in}{polyselection}{A selection of polyhedra: the $\{1\}{\times}\{1\}$, $E_7{\times}SU(2)$~(self-dual) and $E_8{\times}\{1\}$~(self-dual) $K3$ polyhedra. The triangle corresponding to the elliptic fiber divides each polyhedron into a top and bottom.}
\vskip25pt
\framebox[6.5in][c]{\includegraphics[width=5.7in]{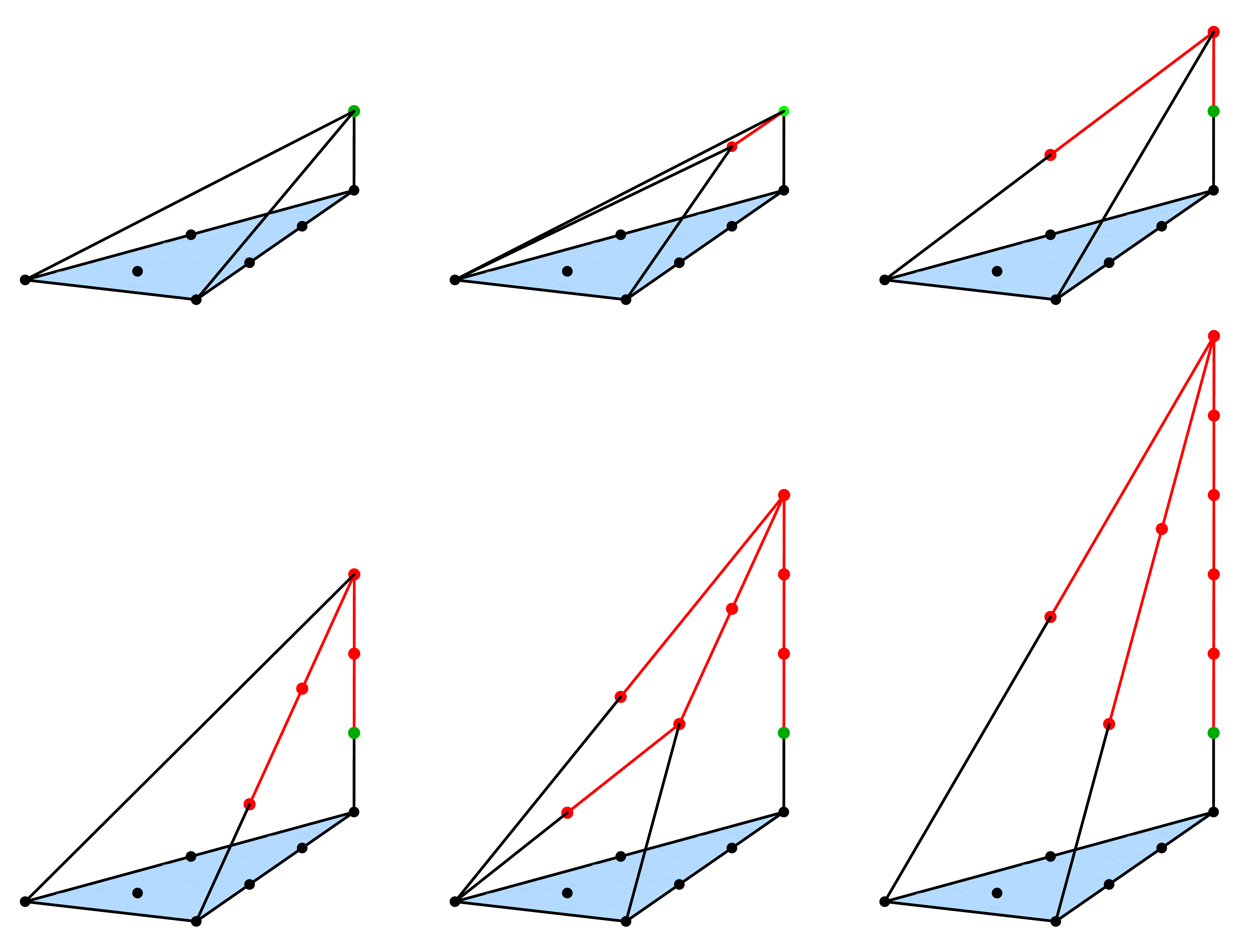}}
\capt{5.2in}{topselection}{A selection of $K3$ tops: the $\{1\}$, $SU(2)$, $G_2$, $F_4$, $E_7$ and $E_8$ tops.}
\end{center}
\end{figure}

The last polyhedron in \fref{polyselection} corresponds to an elliptically fibered $K3$ with a resolved $E_8$ singularity.
In this example, the bottom contains a single lattice point, which gives the Dynkin diagram for the trivial Lie group denoted by $\{1\}$, while the top corresponds to the extended Dynkin diagram for $E_8$.

\subsection{Composition of projecting tops and bottoms}
Originally, a top was defined as a lattice polytope that is `a half of a reflexive polytope' in the sense that it is the part of a reflexive polytope lying on one side of a lattice hyperplane through the origin \cite{Candelas:1996su}.
A more general definition as a lattice polytope with one facet through the  origin and all other facets at distance one from the origin, is useful in the context of non-perturbative gauge groups~\cite{Candelas:1997pq}. All three-dimensional tops of this type were classified in~\cite{Bouchard:2003bu}. In either case the facet of a $d$-dimensional top, that contains the origin, is a $(d-1)$-dimensional reflexive polytope.

In the present paper we are interested in reflexive 4-polytopes
that encode $K3$ fibration structures such that the dual 4-polytope exhibits a 
fibration structure with respect to the dual $K3$ polytope.
This means that in addition to a distinguished lattice vector $m\in M$ 
encoding the slice $\SDelta_3 = \{y\in\SDelta: \langle m,y\rangle=0\}$ 
of $\SDelta$, corresponding to the $K3$ polyhedron,
there is a distinguished lattice vector $n\in N$ encoding the dual slice. 
As shown in~\cite{Avram:1996pj},~$\SDelta$~must then project to $\SDelta_3$
under $\pi_n: N\to N_3\simeq N/N_1$, where $N_1$ is the one-dimensional 
sublattice of $N$ that is generated by $n$.
This implies, of course, that both the top and the bottom resulting from the 
slicing must project to $\SDelta_3$. A top and a bottom that project to~$\SDelta_3$ along the same 
direction $n$ can always be combined into a reflexive polytope, as shown in Lemma~1. In the following we express this fact in terms of a notation where a top is
indicated by a bra~$\Top{A}$, a bottom by a ket $\Bot{B}$
(to be interpreted as the reflection of $\Top{B}$ through $N_3$), and the 
resulting reflexive polytope $\SDelta=\Top{A}\cup\Bot{B}$ by $\topbot{A}{B}$.

\newpage
{\bf Lemma 1.} {\itshape Let $M$, $N$ be a dual pair of four-dimensional lattices
splitting as $N=N_1\oplus N_3$, $M=M_1\oplus M_3$, with $N_1$ generated by 
$n\in N$, $M_1$ generated by $m\in M$ and $\langle m,n\rangle=1$.
Denote by $\pi_n:N\to N_3$, $\pi_m: M\to M_3$ the projections along $n$ and 
$m$, respectively. 
Let $\Delta_3$ and $\SDelta_3$ be a pair of dual polytopes that are reflexive 
with respect to $M_3$ and $N_3$, respectively.
Let $\Top{A}\subset N_\IR$ be a top over $\SDelta_3$, i.e. a lattice polytope 
with one facet inequality $\langle m,y\rangle\ge 0,\, \forall y\in\Top{A}$ such that the facet saturating the inequality is $\SDelta_3$, and all other facet inequalities of the 
type $\langle u_i,y\rangle\ge -1,\,\, \forall y\in \Top{A}$ with $u_i\in M$.
Assume that $\pi_n\Top{A}=\SDelta_3$ and define $\Bot{\widetilde A}\subset M_\IR$ by
$$
\Bot{\widetilde A} = \{x\in M_\IR: ~\langle x,n\rangle\le 0,~
\langle x,y\rangle\ge -1 ~~\forall\,y \in \Top{A} \},
$$
with analogous constructions for the roles of $M$ and $N$ or top and bottom 
reversed. Then:

\begin{itemize}
\item[a)] $\Bot{\widetilde A}$ is a bottom under $\Delta_3$ with $\pi_m\Bot{\widetilde A}=\Delta_3$,

\item[b)] $\Top{\widetilde{\widetilde A}}=\Top{A}$, and

{\item[c)] the union $\topbot{A}{B}$ of $\Top{A}$ and any bottom $\Bot{B}$ with 
$\pi_n\Bot{B}=\SDelta_3$ is a reflexive polytope\\[5pt] with 
$\big(\topbot{A}{B}\big)^{{\!*}}=\topbot{\widetilde B}{\widetilde A}$.}
\end{itemize}
}
{\bf Proof.} 
Let us introduce the semi-infinite prism 
$P^M_{\le }=\IR_{\le }\times \Delta_3\subset M_\IR$ 
(with the subscript `$\le $' meaning $\langle x,n\rangle\le 0$ 
for $x\in P^M_{\le }$) and its analogues
$P^M_{\ge }$, $P^N_{\ge }$ and $P^N_{\le }$.
Then $\Bot{\widetilde A} = \Top{A}^*\cap P^M_{\le }$, where $\Top{A}^*$
is the unbounded polyhedron resulting from applying (\ref{dual}) to $\Top{A}$,
and the projection condition ensures that $\Top{A}^* \supset P^M_{\ge }$,
hence $\Top{A}^* = \Bot{\widetilde A} \cup P^M_{\ge }$. 
\begin{itemize}
\item[{\itshape a)}] The polytope $\Bot{\widetilde A}$ contains $\Delta_3$ as a consequence of its definition 
and the fact that $\Top{A}$ projects to $\SDelta_3$.
The facets of $\Bot{\widetilde A}$ are those of a bottom by definition, and 
its vertices are lattice points since they are either the $u_i$ or vertices of 
$\Delta_3$. 
The bottom $\Bot{\widetilde A} = \Top{A}^*\cap P^M_{\le }$ projects to $\Delta_3$ since 
$P^M_{\le }$ projects to $\Delta_3$.
\item[{\itshape b)}] $\Top{\widetilde {\widetilde A}}=\Bot{\widetilde A}^*\cap P^N_{\ge }
\,=\,\big(\Top{A}\cup (P^M_{\le })^*\big)\cap P^N_{\ge }
\,=\,\big(\Top{A}\cup P^N_{\le }\big)\cap P^N_{\ge }
\,=\,\Top{A}$.
\item[{\itshape c)}] By construction, $\topbot{A}{B}$ is bounded by facets of the type 
$\langle u_i,y\rangle = -1$ and has vertices in~$N$. Convexity is a 
consequence of the projection conditions. The dual is given by
$$\big(\topbot{A}{B}\big)^*=\Top{A}^*\cap \Bot{B}^*
= \big(\Bot{\widetilde A} \cup P^M_{\ge }\big)\cap \big(\Top{\widetilde B}\cup P^M_{\le }\big)=
\topbot{\widetilde B}{\widetilde A}.$$
\hfill$\Box$
\end{itemize}

\subsection{An additivity lemma for the Hodge numbers}
As we saw above, whenever we have projections
both at the $M$ and the $N$ lattice side, the top is determined by the dual 
bottom and vice versa. Also, a top and a bottom that project to $\SDelta_3$ 
along the same direction $n$ can always be combined into a reflexive polytope.
The following lemma shows that, under a specific assumption on the structure
of $\SDelta_3$, this composition obeys additivity in the Hodge numbers of the 
resulting Calabi-Yau threefolds.
\vskip 12pt
{\bf Lemma 2.} {\itshape Let $\SDelta_3$ be a 3-polytope that is reflexive 
with respect to the lattice $N_3$, with no edge $e^*$ 
such that both $e$ and $e^*$ have an interior lattice point.
Let $N=N_1\oplus N_3$, where $N_1$ is generated by the primitive lattice 
vector $n$, and assume that $\Top{A}$ and $\Top{C}$ are tops and $\Bot{B}$ 
and $\Bot{D}$ are bottoms in $N_\IR$ that project to $\SDelta_3$ along $n$. 
Then the relation \eqref{hnosrelation} holds, that is
$$
h^{\bullet\bullet}\big(\topbot{A}{B}\big)+h^{\bullet\bullet}\big(\topbot{C}{D}\big)~=~
h^{\bullet\bullet}\big(\topbot{A}{D}\big)+h^{\bullet\bullet}\big(\topbot{C}{B}\big)~
$$
where $h^{\bullet\bullet}$ stands for the Hodge numbers $\hodgenos$ of the Calabi-Yau 
hypersurface determined by the respective polytope.
}

{\bf Proof.} The Hodge number $h^{1,1}$ 
is given~\cite{Batyrev:1993dm} by
$$ 
h^{1,1}~=~l(\D^*)-5-\hskip-5pt\sum_{{\rm codim\, }\theta^* =1}l^*(\theta^*)+
            \hskip-5pt\sum_{{\rm codim\, }\theta^* =2}l^*(\theta^*)l^*(\theta)~,   
$$
where $l(\SDelta)$ denotes the number of lattice points of $\SDelta$ and 
$l^*(\theta)$ 
denotes the number of interior lattice points of a face $\theta$.

This formula can be rewritten as

$$
h^{1,1} ~= -4 + \hskip-5pt\sum_{P\in\SDelta\cap N} \mathrm{mult}(P)
$$

where the sum runs over all the lattice points $P$ of $\SDelta$ and the 
multiplicities $\mathrm{mult}(P)$ are defined as
$$
\text{mult}(P)~= 
\begin{cases}
0 & \mbox{if } P \mbox{ is the interior point of } \SDelta \mbox{  or is interior to a facet, }\\[4pt] 
1 & \mbox{if } P \mbox{ is a vertex or interior to an edge of } \SDelta, \\[4pt]
\mbox{length}(\theta) & \mbox{if } P \mbox{ is interior to a 2-face } \theta^* \mbox{ of } \SDelta \mbox{ and } \theta \mbox{ is the dual edge of } \Delta, 
\end{cases}
$$
where the length of an edge is the number of integer segments, i.e.~$\text{length}(\theta) = l^*(\theta) +1$.

\vskip10pt
In the following we argue that the contributions of any lattice point $P$ add up to the same value for the different sides of the Hodge number relation. We distinguish the following~cases:
\begin{itemize}
\item[A)] The case $\langle m,P\rangle \ne 0$.
Let us assume, without loss of generality, that $P\in\Top{A}$.
If $P$ is a vertex, or is interior to either a facet or an edge, it will contribute the 
same to $h^{1,1}(\topbot{A}{B})$ as to $h^{1,1}(\topbot{A}{D})$.
Otherwise we use the decomposition $P=\pi_nP+\lambda\, n$ with $\lambda > 0$:
if $P$
is interior to the face $\theta^*$ then any $Q\in \theta$ must satisfy 
$$
-1~=~\langle Q,P\rangle ~=~\langle Q,\pi_nP\rangle+\langle Q,\lambda n\rangle
~\ge~ -1 +\lambda \langle Q,n\rangle~,
$$
hence $\langle Q,n\rangle \le 0$, so all 
of $\theta$ lies in the bottom $\Bot{\widetilde A}$ determined by the top 
$\Top{A}$ to which $P$ belongs.
In other words, 
$P$'s contribution to 
$h^{1,1}(\topbot{A}{B})$ is again the same as its contribution to 
$h^{1,1}(\topbot{A}{D})$.
\item[B)] $P$ is a vertex of $\SDelta_3$. 
Then $P$ is a vertex or interior to an edge
for each of the four polytopes $\topbot{A}{B}, \ldots,$ occurring in 
the Hodge number relation, so $P$ contributes the same value of unity each~time.

\item[C)] $P$ is interior to an edge $e^*$ of $\SDelta_3$.
There are two possibilities:
\begin{itemize}
\item[a)] $e^*$ is an edge of $\SDelta$, in which case $P$ contributes 1 to  
$h^{1,1}$.
\item[b)] $P$ lies within a two-face of $\SDelta$ which is dual 
(in the four-dimensional sense) 
to \hbox{$e\subset \Delta_3\subset\Delta$}. 
By our assumptions, $e$ has length 1 so $P$ again~contributes~1.
\end{itemize}
\item[D)] $P$ is interior to a facet of $\SDelta_3$ that is dual to a vertex 
$v$ of $\Delta_3$. In $\SDelta$, the face $\theta^*$ to which $P$ is interior 
can be a facet dual to $v$, implying $\text{mult}(P)=0$, or a codimension 2-face dual to an edge $\theta$ that projects to $v$, in which case 
$\text{mult}(P)=\text{length}(\theta)$.
The length of such an edge is additive under the composition of tops with 
bottoms, hence the contribution of $P$ is additive again.
\end{itemize}

This shows additivity for $h^{1,1}$. 
Additivity for $h^{1,2}$ follows from the compatibility of top-bottom 
composition with mirror symmetry for projecting tops and bottoms.
\hfill$\Box$
 
\vskip 8pt
The assumption on the dual pairs of edges of the $K3$ polytopes is
necessary, as the following example shows. Consider the polytope associated with the gauge group 
$F_4{\times} G_2$, as in the following figure. This polyhedron is self-dual and possesses dual pairs of edges of length $>1$.

\begin{figure}[bt]
\begin{center}
\framebox{\hskip30pt
\begin{minipage}[t]{4.5in}
 $\begin{array}{rrrr}
(&\hskip-10pt -2,&2, &~3)\\
(&                   0,&-1, &0)\\
(&                   0,&  0,&-1)\\
(&                   3,&  2,&3)\end{array}$
\hfill 
\raisebox{-1in}{\includegraphics[width=5.cm]{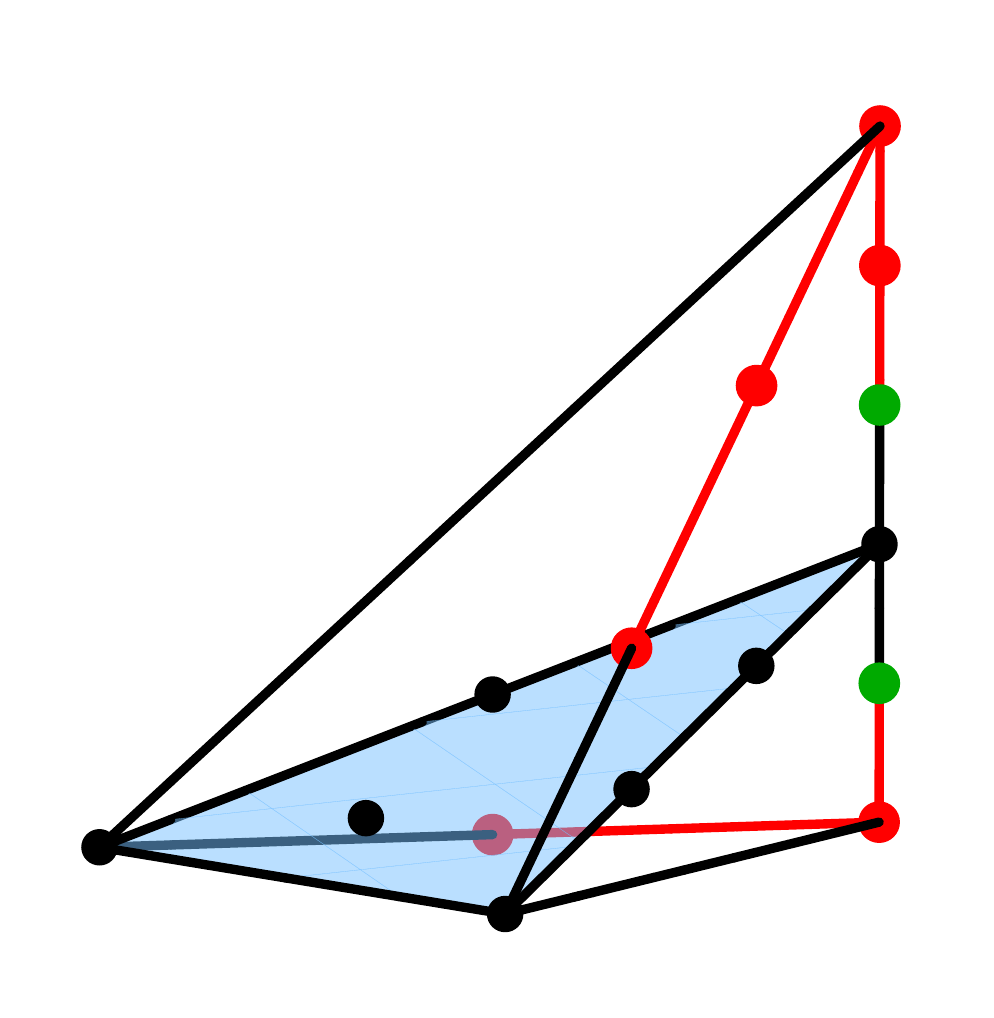}}
\hspace*{25pt}
\end{minipage}}
\capt{6.3in}{F4G2poly}{The polyhedron for an elliptically fibered $K3$ manifold of the type $F_4{\times} G_2$.}
\end{center}
\vskip -10pt
\end{figure}

A minimal extension of the $F_4{\times}G_2$ polyhedron to a four-dimensional top
\[
\Top{\text{min}} = \text{Conv}\Big(\big\{(0, -2, 2, 3),\, (0, 0, -1, 0),\, (0, 0,  0, -1),\, (0, 3, 2, 3),\, 
(1, 0, 0, 0)\big\}\Big)
\]
is easily seen to be dual to a prism shaped bottom
$\Bot{\text{prism}} = [-1,0]\times\SDelta_3$.

One finds the Hodge numbers 
\begin{equation*} 
\big(h^{1,1},h^{1,2} \big)\big(\topbot{\text{min}}{\text{min}}\big)=(21,45),~~~
\big(h^{1,1},h^{1,2} \big)\big(\topbot{\text{prism}}{\text{prism}}\big)=(45,21),
\end{equation*}
while
\begin{equation*}
\big(h^{1,1},h^{1,2} \big)\big(\topbot{\text{min}}{\text{prism}}\big)=
\big(h^{1,1},h^{1,2} \big)\big(\topbot{\text{prism}}{\text{min}}\big)=(31,31)
\end{equation*}
and we see that these Hodge numbers violate Formula the Hodge number relation.

For constructing this example we have used the fact that an edge in a toric 
diagram corresponding to a Lie group satisfies the assumption precisely if the 
Lie group is simply laced, i.e.~of ADE type~\cite{Perevalov:1997vw}. 
However, for applying Lemma 2 we do not necessarily require an elliptic 
fibration; conversely, an elliptic fibration structure with simply laced gauge 
groups is not sufficient since edges of the fiber polygon may 
violate the condition.
\newpage
\section{The web of $E_8{\times} \{1\}$ elliptic $K3$ fibrations}\label{E81}
The special structures which served as motivating examples at the beginning of this paper, together with the observation that some of the points in these structures correspond to $K3$ fibered Calabi-Yau manifolds of the type presented above, lead us to the problem of searching for $E_8{\times} \{1\}$ $K3$ fibrations through the list of reflexive 4-polytopes. This search is described~below. 

The polytope $\SDelta_{E_8{\times}\{1\}}$ is
isomorphic to the convex hull of the vertices
$$
\begin{array}{rrrr} 
(&\hskip-10pt -1,& 2,& 3)\\[2pt]
(&                   0,& -1,& 0)\\[2pt]
(&                   0,& 0, &-1)\\[2 pt]
(&                   6,& 2, & 3)\rlap{~.}
\end{array}
$$
The polyhedron is shown in \fref{polyselection}. By extending the lattice into a fourth dimension and adding the point $(1, 6, 2, 3)$, one obtains a top that we denote by $\Top{\text{min}}$. Similarly, adding the point $(-1, 6, 2, 3)$ results in the bottom $\Bot{\text{min}}$.
Adding both points gives the reflexive polytope $\topbot{\text{min}}{\text{min}}$, whose vertices are listed in Table \ref{E8SU1vertices}, along with the vertices for the dual~polytope $\topbot{\text{max}}{\text{max}}$.

\begin{table}[htdp]
\def\str{\varstr{12pt}{6pt}}
\begin{center}
\begin{tabular}{| >{$~~} r <{~~$} | >{$~~} r <{~~$} | >{$~~} r <{~~$} |}
\hline
\varstr{16pt}{8pt} \topbot{\text{min}}{\text{min}}^{11,491}
& \topbot{\text{max}}{\text{max}}^{491,11} & \topbot{\text{min}}{\text{max}}^{251,251} \\
\hline\hline
\varstr{14pt}{6pt} (-1,\ \ \ 6,\ \ \ 2,\ \ \ 3) & (-42,\ \ \ 6,\ \ \ 2,\ \ \ 3) & (-42,\ \ \ 6,\ \ \ 2,\ \ \ 3) \\
\str (\ \ 0,\ -1,\ \ \ 2,\ \ \ 3)     & (\hskip14pt 0,\  -1,\ \ \ 2,\ \ \ 3)       & (\hskip14pt 0,\ -1,\ \ \ 2,\ \ \ 3) \\
\str (\ \ 0,\ \ \ 0,\  -1,\ \ \ 0)    & (\hskip14pt 0,\ \ \ 0,\  -1,\ \ \ 0)       & (\hskip14pt 0,\ \ \ 0,\  -1,\ \ \ 0) \\
\str (\ \ 0,\ \ \ 0,\ \ \ 0,\  -1)    & (\hskip14pt 0,\ \ \ 0,\ \ \ 0,\  -1)       & (\hskip14pt 0,\ \ \ 0,\ \ \ 0,\  -1) \\
\varstr{12pt}{8pt} 
(\hskip9pt 1,\ \ \ 6,\ \ \ 2,\ \ \ 3)  & (\hskip9.5pt 42,\ \ \ 6,\ \ \ 2,\ \ \ 3)   & (\hskip16pt 1,\ \ \ 6,\ \ \ 2,\ \ \ 3) \\
\hline \hline
\end{tabular}
\capt{5.0in}{E8SU1vertices}{The minimal, the maximal and the maximal self-dual reflexive polytopes containing the $E_8{\times}\{1\}$ $K3$~polyhedron as a slice and as a projection.}
\end{center}
\end{table}
The Calabi-Yau threefold $\cM_{11,491}$ (the subscripts refer to its Hodge numbers) is determined by $\SDelta_{11,491}\cong\topbot{\text{min}}{\text{min}}$ and $\Delta_{11,491}\cong\topbot{\text{max}}{\text{max}}$. Combining $\Top{\text{min}}$ with $\Bot{\text{max}}$
results in a self-dual reflexive polytope $\topbot{\text{min}}{\text{max}}$ whose vertices are listed in the third column of Table~\ref{E8SU1vertices}, corresponding to the self-mirror threefold $\cM_{251,251}$. This manifold with vanishing Euler number is indicated in \fref{BasicKSPlot} by the topmost point lying on the axis 
$\chi=0$. 

The manifolds $\cM_{11,491}$ and $\cM_{251,251}$ are related by what we called `half-mirror symmetry' in the introduction. In fact, as discussed below, `half-mirror symmetry' corresponds to a combination of mirror symmetry with replacing $\Top{\text{max}}$ by $\Top{\text{min}}$ or $\Bot{\text{max}}$ by 
$\Bot{\text{min}}$.

\subsection{Searching for $E_8{\times} \{ 1\}$ elliptic $K3$ fibrations}
Adding points to the polytope $\SDelta_{11,491}\simeq\topbot{\text{min}}{\text{min}}$  
will not decrease $h^{1,1}$. This can be seen as follows.  
For the smooth fibration $\cM_{11,491}$, the Picard number $h^{1,1} = 11$ comes from the 10 toric divisors in the $K3$, as well as from the generic fiber (the $K3$ itself). Enlarging the top/bottom corresponds to blowing up the points $z = 0$ or $z = \infty$ of the $\mathbb P^1$ that is the base of the fibration. These blowups take place separately at the two distinguished points and add up to 240 exceptional divisors at each of them. Denote by $N_{\text{top}}$ and $N_{\text{bottom}}$ the number of exceptional divisors resulting from adding points in the top 
$\Top{\text{min}}$ and bottom $\Bot{\text{min}}$ respectively. Then we have
$h^{1,1} = 11 + N_{\text{bottom}} + N_{\text{top}}$ with 
$0\leq N_{\text{top}}, N_{\text{bottom}}\leq 240$. The maximal bottom corresponds to 
$h^{1,1} = 251 + N_{\text{top}}$. 

This argument tells us that the reflexive polytopes containing the maximal bottom and an arbitrary top are characterised by $h^{1,1}\geq 251$ and positive Euler number. The list contains 2,219 polytopes that pass both of these requirements. We identify the polytopes that contain a maximal bottom by searching for a distinguished 3-face containing 4 vertices and 24 points. In the representation 
of \tref{E8SU1vertices}
the facet in question has vertices
$$ 
\big\{(-42, 6, 2, 3),\ (0, -1, 2, 3),\ (0, 0, -1, 0),\ (0, 0, 0, -1)\big\}~.
$$
It is important to note that this facet is not orthogonal to the hyperplane determined by the polyhedron associated with the $K3$. As such, since we are searching for polytopes which contain the $K3$ polyhedron both as a slice and as a projection (corresponding to $K3$ fibrations for which the mirror image is also $K3$ fibration), this facet cannot extend into the top half. This means that, in order to find all reflexive polytopes containing the maximal bottom, it is enough to search for those reflexive polytopes which contain the distinguished facet and then check that this facet indeed belongs to a maximal bottom. 

The search yields a list of 1,263 reflexive polytopes, and thus an equal number of distinct tops, which are available at the URL given in \SS\ref{webaddress}. The Hodge numbers associated with these reflexive polytopes (465 distinct Hodge pairs) are shown in red in \fref{BigBotPoints}.  

\subsection{Generating all $E_8{\times} \{1\}$ elliptic $K3$ fibrations} 
It is now easy to generate a full list of polytopes corresponding to Calabi-Yau threefolds exhibiting the $E_8{\times} \{1\}$ $K3$ fibration structure discussed above. Indeed, this can be realised by taking all possible combinations $\topbot{A}{B}$, with $\Top{A}$ and $\Bot{B}$ being tops and bottoms from the previous list, glued along the polyhedron corresponding to the $K3$ fiber. 
There are 798,216 such reflexive polytopes. The 16,148 distinct Hodge pairs associated with these polytopes are shown in red in \fref{AllE8TimesSU1}. Prior to having a proof of \eqref{hnosrelation} we checked the following, equivalent, relation for each of the combinations:
$$
h^{\bullet\bullet}\big(\topbot{A}{B}\big)~=~ h^{\bullet\bullet}\big(\topbot{A}{\text{max}}\big)+
h^{\bullet\bullet}\big(\topbot{\text{max}}{B}\big)-h^{\bullet\bullet}\big(\topbot{\text{max}}{\text{max}}\big)~.
$$
\newpage
\section{More Webs}\label{MoreWebs}
\subsection{The web of $E_7{\times} SU(2)$ elliptic $K3$-fibrations}\label{E71}

For the elliptic $K3$ surface with degenerate fibers of the type $E_7 {\times} SU(2)$ we consider the polyhedron given by the vertices shown in \fref{K3polyhedron2}. As before, we extend this polyhedron to a 4-dimensional reflexive polytope by adding two points, above and below the point $(0,4,2,3)$. The resulting polytope, as well as its dual are given in~Table~\ref{E7SU2vertices}.
\vskip20pt
\begin{figure}[H]
\begin{center}
\framebox{\hskip30pt
\begin{minipage}[t]{4.5in}
$\begin{array}{rrrr}
(&\hskip-10pt -1,& 1,& 2)\\
(&\hskip-10pt -1,& 2,& 3)\\
(&                   0,& -1,& 0)\\
(&                   0,&  0,&-1)\\
(&                   0,&  2,& 3)\\
(&                   2,&  0,& 1)\\
(&                   4,&  2,& 3)\end{array}$
\hfill
\raisebox{-1.1in}{ \includegraphics[width=5.cm]{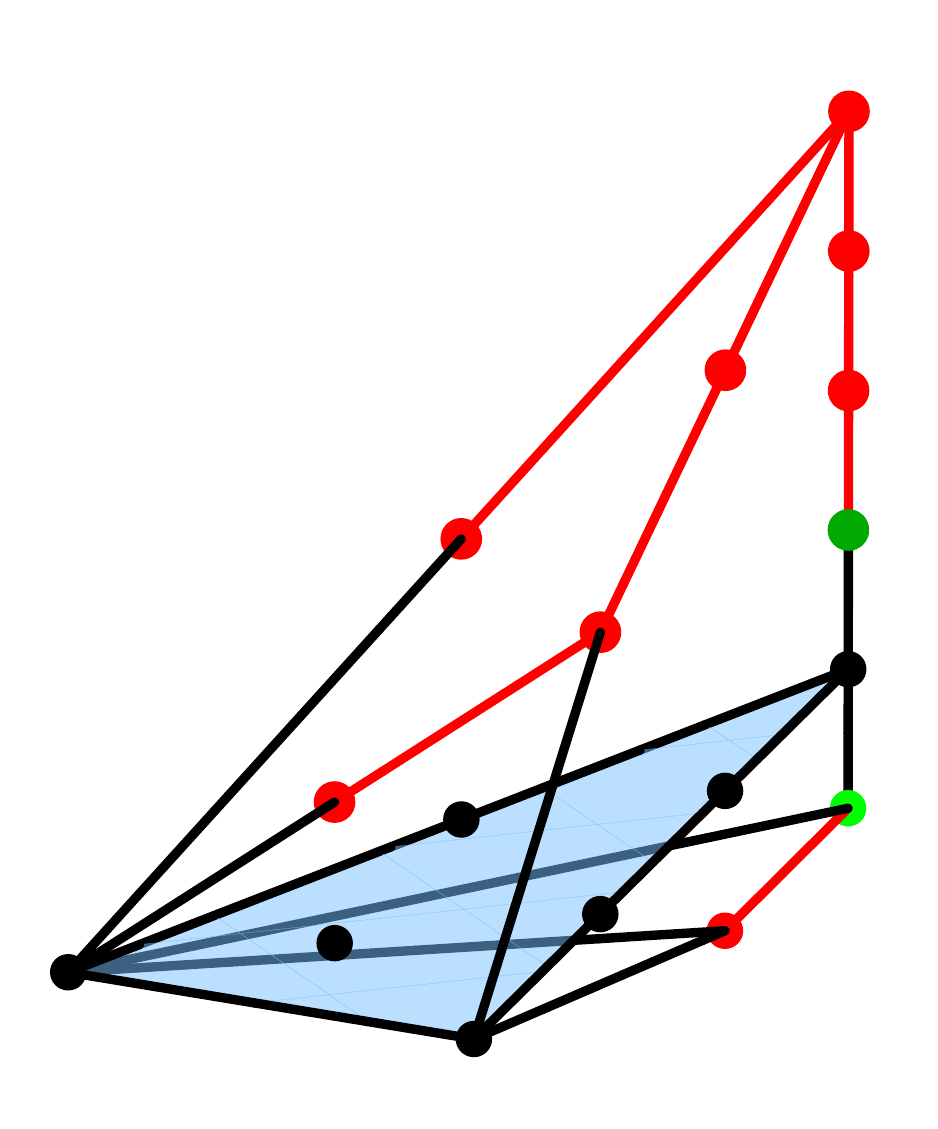}}
\hspace*{25pt}
\end{minipage}}
\capt{6.3in}{K3polyhedron2}{Elliptically fibered $K3$ manifold of the type $E_7{\times} SU(2)$.}
\end{center}
\end{figure}
\vfill
\begin{table}[H]
\def\str{\varstr{12pt}{6pt}}
\begin{center}
\begin{tabular}{| >{$~~} r <{~~$} | >{$~~} r <{~~$} |}
\hline
\varstr{16pt}{8pt} 
\topbot{\text{min}}{\text{min}}^{11,251} & \topbot{\text{max}}{\text{max}}^{251,11} \\
\hline\hline
\varstr{14pt}{6pt} (-1,\ \ \ 4,\ \ \ 2,\ \ \ 3) & (-22,\ \ \ 4,\ \ \ 2,\ \ \ 3) \\
\str (\ \ 0,\ -1,\ \ \ 1,\ \ \ 2) & (\hskip13.5pt 0,\ -1,\ \ \ 1,\ \ \ 2) \\
\str (\ \ 0,\ -1,\ \ \ 2,\ \ \ 3) & (\hskip13.5pt 0,\ -1,\ \ \ 2,\ \ \ 3) \\
\str (\ \ 0,\ \ \ 0,\ -1,\ \ \ 0) & (\hskip13.5pt 0,\ \ \ 0,\ -1,\ \ \ 0) \\
\str (\ \ 0,\ \ \ 0,\ \ \ 0,\  -1) & (\hskip13.5pt 0,\ \ \ 0,\ \ \ 0,\  -1) \\
\str (\ \ 0,\ \ \ 0,\ \ \ 2,\ \ \ 3) & (\hskip13.5pt 0,\ \ \ 0,\ \ \ 2,\ \ \ 3) \\
\str (\ \ 0,\ \ \ 2,\ \ \ 0,\ \ \ 1) & (\hskip13.5pt 0,\ \ \ 2,\ \ \ 0,\ \ \ 1) \\
\varstr{12pt}{8pt} (\hskip7.5pt 1,\ \ \ 4,\ \ \ 2,\ \ \ 3) & (\hskip8pt 22,\ \ \ 4,\ \ \ 2,\ \ \ 3) \\
\hline \hline
\end{tabular}
\capt{4.5in}{E7SU2vertices}{The minimal and the maximal reflexive polytopes containing the $E_7\times SU(2)$ $K3$~polyhedron as a slice and as a projection.}
\end{center}
\end{table}

As before, we are interested in finding all the reflexive polytopes which contain the $K3$ polyhedron of type $E_7{\times} SU(2)$ both as a slice and as a projection. The search for all the tops and bottoms that can be joined with the $K3$ polyhedron, performed in a way similar to the $E_8{\times} \{1\}$ case results in a list of $1204$ different tops and so also in $1204$ different bottoms. As in the previous case the set of polytopes
$\topbot{A}{\text{max}}$, as $\Top{A}$ ranges over all tops, gives rise to a $V$-structure, shown as the region formed by the red points on the left of \fref{AllE7TimesSU2} that is bounded by the blue lines. Within this region, the polytopes of greatest height are $\D=\topbot{\text{max}}{\text{max}}^{11,251}$ and 
$\D=\topbot{\text{min}}{\text{max}}^{131,131}$. The analogue of the vector that was 
$\D\hodgenos=(1,-29)$, for the case $E_8{\times}\{1\}$, is now $\D\hodgenos=(0,-16)$ and this vector, together with its mirror, determine the slope of the bounding lines. The analogue of the vector, that was previously, $\D\hodgenos=(240,-240)$ is now $\D\hodgenos=(120,-120)$.
\begin{figure}[!t]
\begin{center}
\includegraphics[width=6.5in]{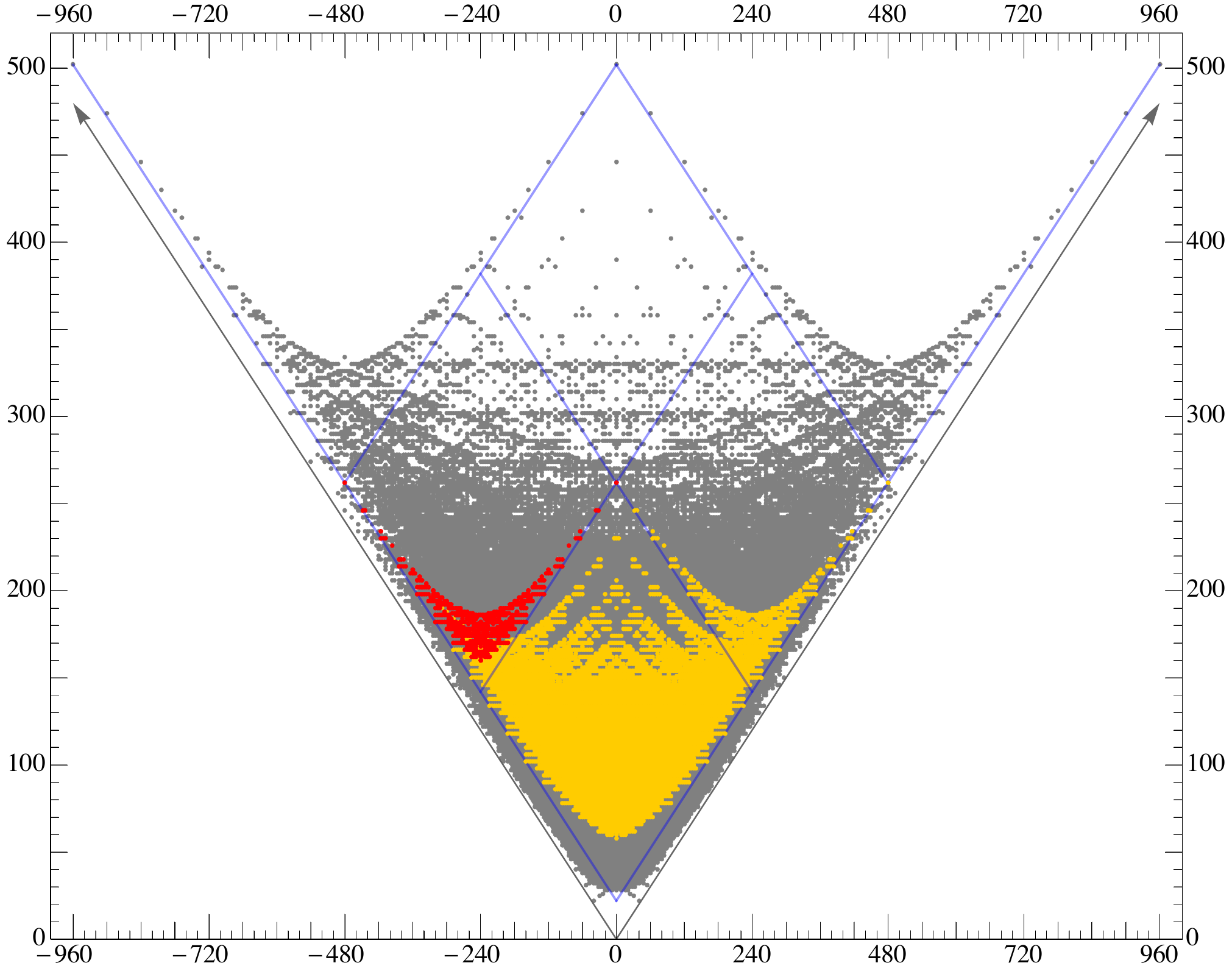}
\capt{6.2in}{AllE7TimesSU2}{Hodge plot showing all $E_7{\times}SU(2)$ $K3$ fibrations for which the fiber is both a slice and a projection in the four-dimensional reflexive polytope. The points in red form the $V$-region, corresponding to taking, for $\D$, all possible tops together with the maximal bottom.}
\end{center}
\vskip-20pt
\end{figure}

By combining tops and bottoms from these lists, we obtain a number of $725,410$ polytopes which correspond to Calabi-Yau manifolds which are elliptic $K3$ fibrations of the type $E_7{\times} SU(2)$. Associated with these manifolds, there are $7,929$ distinct Hodge number pairs, indicated by the coloured points in \fref{AllE7TimesSU2}. The additive property for the Hodge numbers holds also in this case. It is interesting to note the similarity between the plots in \fref{AllE8TimesSU1} and \fref{AllE7TimesSU2}. The particular shape of the structures present in these plots seems to be a generic feature of the webs of elliptic $K3$ fibered Calabi-Yau manifolds with a self-dual $K3$ manifold. 

\vskip-5pt
\goodbreak
\subsection{The web of $E_7{\times} \{1\}$ and $E_8{\times} SU(2)$ elliptic $K3$-fibrations} 
\vskip-4pt
In the case when the $K3$ manifold is not self-dual, one needs to consider two webs at a time. For example, an elliptic $K3$ fibration of the type $E_7{\times} \{1\}$ is dual to an elliptic $K3$ fibration of the type $E_8{\times} SU(2)$. These polyhedra appear in \fref{K3polyhedron3}. Their vertices are listed in Table~\ref{E7SU1andE8SU2K3vertices}. 
\vskip-1pt
\begin{figure}[H]
\begin{center}
\framebox[5.8in][c]{
$\begin{array}{cc}
\hspace{-30pt}  
\includegraphics[width=3.7cm]{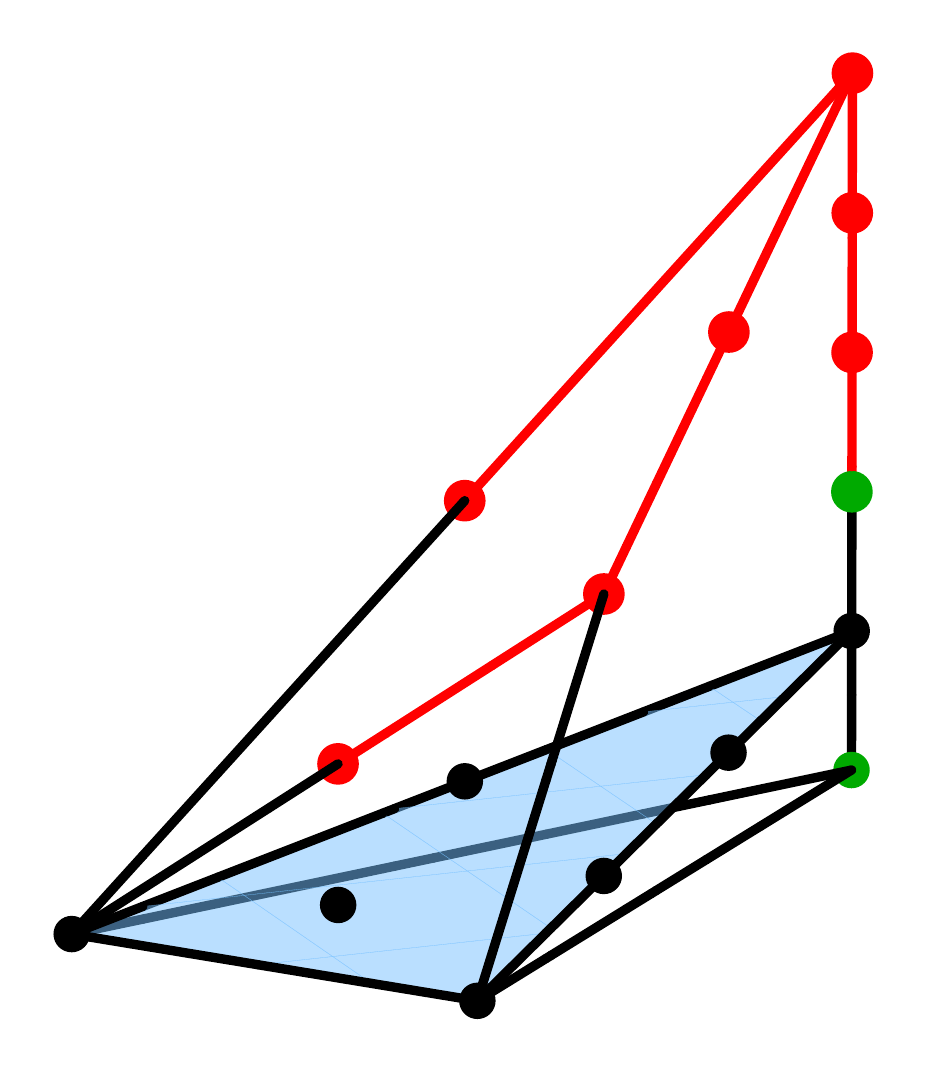}&\hspace{50pt}
\includegraphics[width=3.7cm]{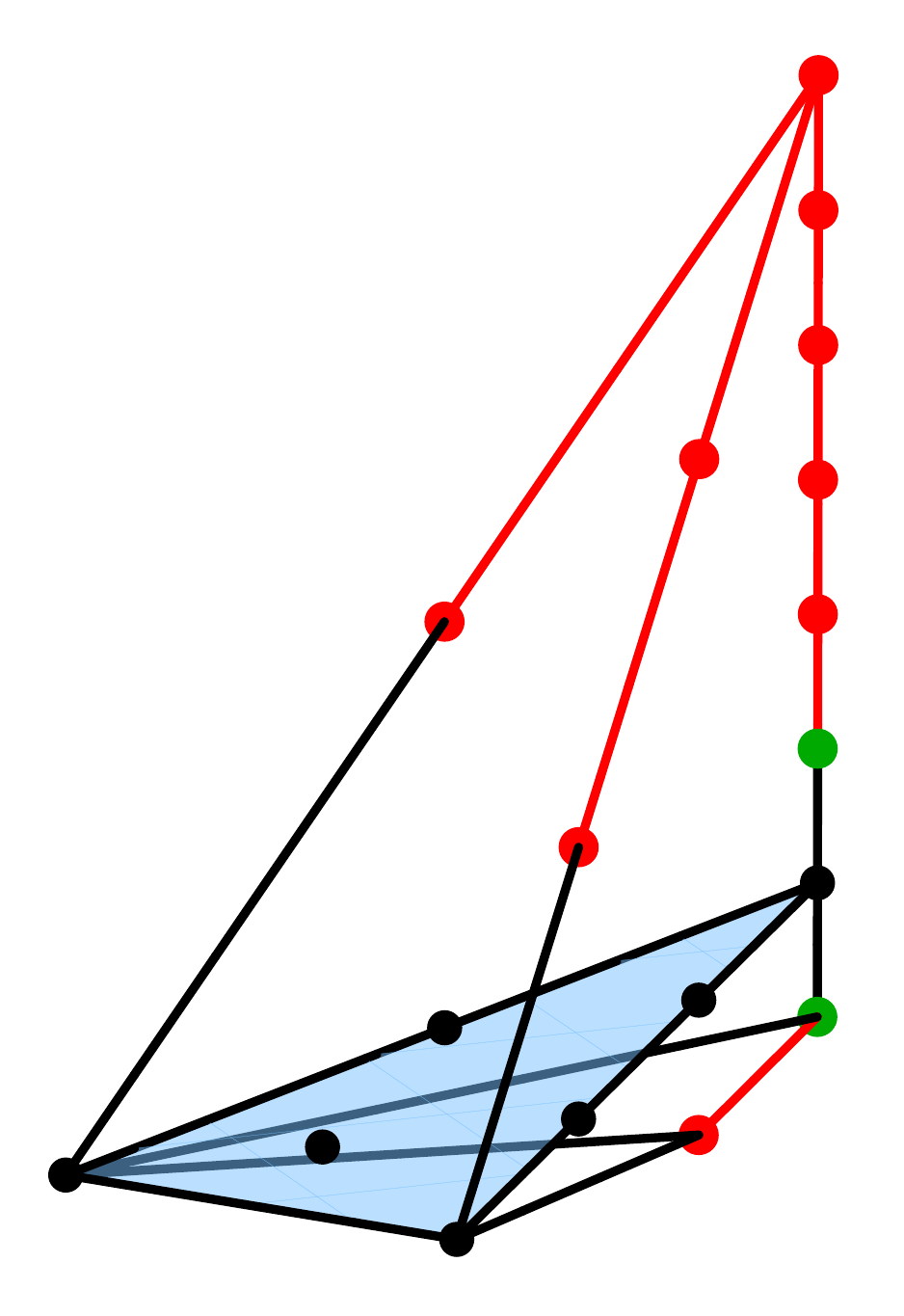}
\end{array}$
}
\capt{6.5in}{K3polyhedron3}{Elliptically fibered $K3$ manifolds of type $E_7{\times} \{1\}$ (left) and of type $E_8{\times} SU(2)$ (right).}
\end{center}
\vskip-20pt
\end{figure}

The minimal and maximal 4D reflexive extensions of the $E_7{\times} \{1\}$ and the $E_8{\times} SU(2)$ $K3$~polyhedra are listed in Tables \ref{E7SU1vertices}.
The combined web of $E_7{\times} \{1\}$ and $E_8{\times} SU(2)$  contains $14,356$ distinct Hodge number pairs. This web is indicated by the coloured structure in \fref{E7E8str}. The red and the blue points correspond to $K3$ fibrations of the $E_7\times \{1\}$ and $E_8\times SU(2)$ type, respectively. The purple points correspond to fibrations of both types. Note the similarity between the structure formed by the purple points and the previous webs associated to self-dual $K3$ polyhedra.
\begin{table}[H]
\def\str{\varstr{11pt}{5pt}}
\begin{center}
\begin{tabular}{| >{$~~~} r <{~~~$} | >{$~~~} r <{~~~$} |}
\hline
\varstr{15pt}{9pt}
\big(\SDelta_3 \big)_{E_7\times\{1\}} & \big(\SDelta_3 \big)_{E_8\times SU(2)} \\
\hline\hline
\varstr{14pt}{6pt} (-1,\ \ \ 2,\ \ \ 3) & (-1,\ \ \ 1,\ \ \ 2) \\
\str (\ \ 0,\ -1,\ \ \ 0) & (-1,\ \ \ 2,\ \ \ 3) \\
\str (\ \ 0,\ \ \ 0,\  -1) & (\ \ 0,\ -1,\ \ \ 0) \\
\str (\ \ 2,\ \ \ 0,\ \ \ 1) & (\ \ 0,\ \ \ 0,\  -1) \\
\varstr{12pt}{8pt} (\ \ 4,\ \ \ 2,\ \ \ 3) & (\ \ 6,\ \ \ 2,\ \ \ 3) \\
\hline \hline
\end{tabular}
\capt{5.9in}{E7SU1andE8SU2K3vertices}{The $E_7\times\{1\}$ and $E_8\times SU(2)$ $K3$~polyhedra.}
\end{center}
\vskip-10pt
\end{table}
\begin{figure}[H]
\begin{center}
\includegraphics[width=6.5in]{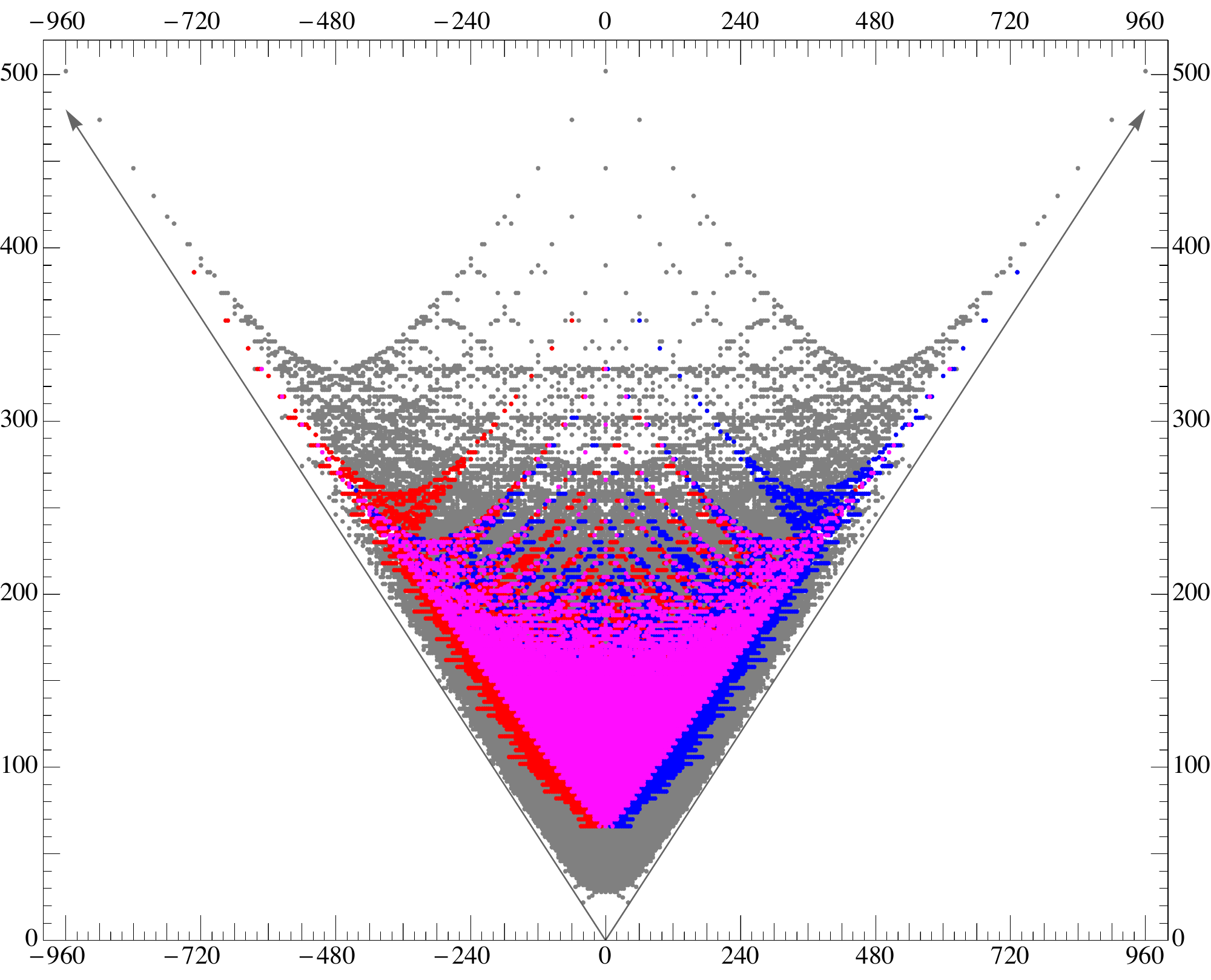}
\capt{5.0in}{E7E8str}{Hodge plot showing the $E_7{\times} \{1\}$ and $E_8{\times} SU(2)$ web.}
\end{center}
\end{figure}
\begin{table}[htdp]
\def\str{\varstr{12pt}{6pt}}
\begin{center}
\begin{tabular}{| >{$~} r <{~$} | >{$~} r <{~$} |  >{$~} r <{~$} |  >{$~} r <{~$} |}
\hline
\varstr{19pt}{12pt}
\topbot{\text{min}}{\text{min}}^{10,376}_{E_7\times \{1\}} 
& \topbot{\text{max}}{\text{max}}^{318,12}_{E_7\times \{1\}}
  & \topbot{\text{min}}{\text{min}}^{12,318}_{E_8{\times} SU(2)} 
     & \topbot{\text{max}}{\text{max}}^{376,10}_{E_8{\times} SU(2)} \\
\hline\hline
\varstr{14pt}{6pt} (-1,\ \ \ 4,\ \ \ 2,\ \ \ 3) & (-30,\ \ \ 4,\ \ \ 2,\ \ \ 3) 
                 &(-1,\ \ \ 6,\ \ \ 2,\ \ \ 3) & (-30,\ \ \ 6,\ \ \ 2,\ \ \ 3) \\
\str (\ \ 0,\ -1,\ \ \ 2,\ \ \ 3) & (-14,\ \ \ 2,\ \ \ 0,\ \ \ 1) 
                 & (\ \ 0,\ -1,\ \ \ 1,\ \ \ 2) & (\hskip4.5pt -2,\ -1,\ \ \ 2,\ \ \ 3) \\
\str (\ \ 0,\ \ \ 0,\ -1,\ \ \ 0) & (\hskip13.5pt 0,\ -1,\ \ \ 2,\ \ \ 3) 
                 & (\ \ 0,\ -1,\ \ \ 2,\ \ \ 3)  &  (\hskip13.5pt 0,\ -1,\ \ \ 1,\ \ \ 2) \\
\str (\ \ 0,\ \ \ 0,\ \ \ 0,\  -1)  &  (\hskip13.5pt 0,\ \ \ 0,\ -1,\ \ \ 0) 
                 & (\ \ 0,\ \ \ 0,\ -1,\ \ \ 0)  & (\hskip13.5pt 0,\ \ \ 0,\ -1,\ \ \ 0) \\
\str (\ \ 0,\ \ \ 2,\ \ \ 0,\ \ \ 1) & (\hskip13.5pt 0,\ \ \ 0,\ \ \ 0,\  -1) 
                 & (\ \ 0,\ \ \ 0,\ \ \ 0,\  -1) & (\hskip13.5pt 0,\ \ \ 0,\ \ \ 0,\  -1) \\ 
\str (\ \ 1,\ \ \ 4,\ \ \ 2,\ \ \ 3) & (\hskip8pt 14,\ \ \ 2,\ \ \ 0,\ \ \ 1) 
                 & (\hskip9.5pt 1,\ \ \ 6,\ \ \ 2,\ \ \ 3) & (\hskip13.5pt 2,\ -1,\ \ \ 2,\ \ \ 3) \\
\varstr{12pt}{8pt} & (\hskip8pt 30,\ \ \ 4,\ \ \ 2,\ \ \ 3)&& (\hskip9.5pt 30,\ \ \ 6,\ \ \ 2,\ \ \ 3) \\
\hline \hline
\end{tabular}
\capt{6in}{E7SU1vertices}{Minimal and maximal reflexive polytopes, $\D^*$, with a $K3$ slice of type $E_7{\times}\{1\}$  (first two columns) or an $E_8\times SU(2)$ slice (last two columns).}
\end{center}
\vspace{-20pt}
\end{table}
\newpage
\section{Outlook}\label{Outlook}
We have seen that
the intricate structure of the upper region of the Hodge plot associated with the 
list of reflexive 4-polytopes can be largely understood as an overlap of webs of $K3$ fibrations. The pattern formed by the points of each web resembles that of the entire plot. Although very intricate, this pattern has a very regular structure, being formed by replicating a certain substructure many times. These intricately self-similar nested patterns within patterns give to the Hodge plot the appearance of a fractal. 
\begin{figure}[!t]
\begin{center}
\includegraphics[width=6.2in]{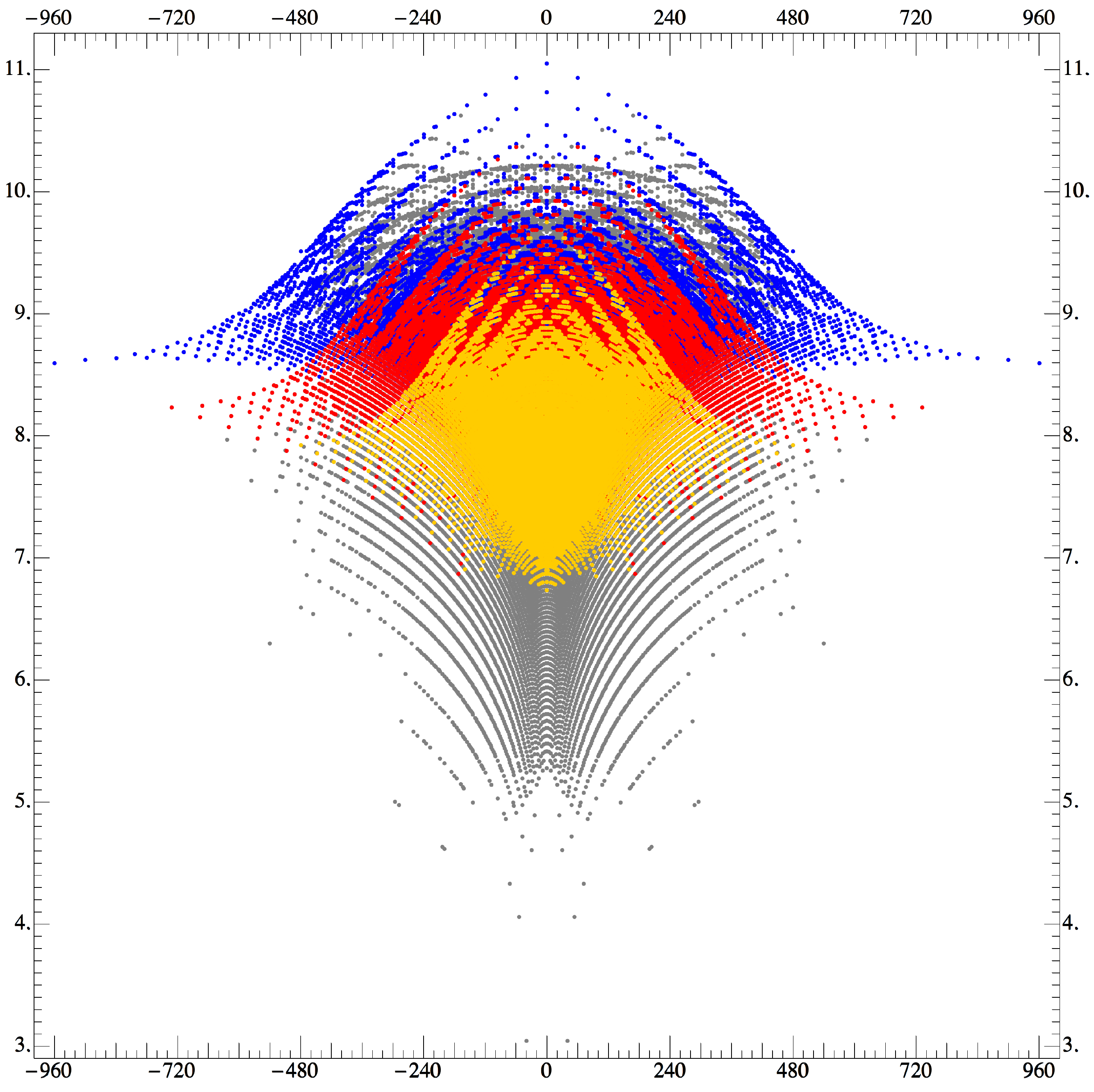}
\capt{5.5in}{NewKSPlot2}{Hodge plot showing in blue all $E_8\times \{1\}$ $K3$ fibrations; in red all $E_7\times \{1\}$ and $E_8\times SU(2)$ $K3$ fibrations; in green: all $E_7{\times}SU(2)$ $K3$ fibrations. Horizontal axis: the Euler number. Vertical axis: $\log h^{1,1}{+}\log h^{1,2}$.}
\end{center}
\vskip-10pt
\end{figure}
The plot contains many such webs, according to the different types of $K3$ manifolds. In this paper we have considered only three. Despite making this very restricted choice, we obtain a great number of $K3$ fibrations corresponding to $20,947$ distinct Hodge pairs, out of a total of $30,108$ Hodge~pairs. 

The plots in \fref{RBGplot} and \fref{NewKSPlot2} display the three overlapping webs discussed in this paper. The second of these plots uses a different coordinate system, $\log h^{1,1}+\log h^{1,2}$ against the Euler number. The most striking feature of this plot is the fact that the webs seem to~separate.  

How could one proceed with a study of the type presented here?
On the one hand, one can continue to study the Hodge plot and try to identify 
manifolds corresponding to Hodge numbers that do not belong to one of the webs 
identified so far.
In this way we found the $E_7{\times} \{1\}$ web as the structure providing the 
first (gray) points in the top left corner of \fref{AllE8TimesSU1} that 
cannot be explained by the $E_8{\times} \{1\}$ web (red points).
Similarly, the first gray point in the top left corner of \fref{RBGplot}
corresponds to $h^{1,1}=9$, $h^{1,2}=321$.
There are three polytopes $\SDelta$ giving rise to these Hodge numbers.
All of them are fibrations of the $\topbot{\text{min}}{\text{min}}$
type over elliptic $K3$ manifolds, with the $K3$ polytopes $\SDelta_3$ being of the types 
$F_4{\times} \{1\}$, $E_6{\times} \{1\}$ and $E_6'{\times} \{1\}$, respectively;
the 3-dimensional tops of $E_6$ and $E_6'$ type both correspond to the Lie group $E_6$, 
but the $E_6'$ diagram has one lattice point more than the $E_6$ diagram.

What about the gray dots remaining in the upper central portion of 
\fref{RBGplot}?
It turns out that at least the first few of these correspond to cases where
either the manifold or its mirror is a fibration over an elliptic $K3$ of the 
$E_8{\times} \{1\}$ type; remember that for the models of \sref{E81} both
the manifold and its mirror were of this type.
In other words, now either slicing or projecting gives the last polyhedron of 
\fref{polyselection}, but the $E_8\times\{1\}$ polyhedron is not both a slice and a projection.

A different approach to identifying structures in the set of toric Calabi-Yau 
hypersurfaces is to work directly with the polytopes.
The classification of all reflexive 4-polytopes used the fact that there is a 
set of only 308 maximal polytopes (listed in Appendix A of 
\cite{Kreuzer:2000xy}) that contain all reflexive polytopes as subpolytopes, 
possibly on sublattices.
Their duals are minimal polytopes in the sense that any reflexive polytope 
can be obtained from a minimal one by adding lattice points.
This means that every Calabi-Yau hypersurface can be obtained via blowing up 
(and, possibly, orbifolding)
a Calabi-Yau hypersurface corresponding to a pair of polytopes with $\SDelta$ 
minimal, $\Delta$ maximal.

It turns out that most of the minimal polytopes exhibit nested fibration
structures, typically with Weierstrass elliptic fibers.
There are 40 minimal polytopes giving rise to Euler numbers $\ch \le -480$.
The corresponding list of 40 maximal polytopes contains all the maximal 
polytopes with 340 or more points (half of the maximal point number of 680).
All of these polytopes correspond to Weierstrass models, which fits nicely
with recent observations~\cite{Taylor:2012dr} about the connections between
such fibrations and the structure of the Hodge plot.
Of the 40 models, 38 are also $K3$ fibered, with the Weierstrass triangle 
slicing the
reflexive 3-polytope $\SDelta_3$ encoding the $K3$; the extension to the
reflexive 4-polytope $\SDelta$
is always of the $\topbot{\text{min}}{\text{min}}$ type, i.e.~with just one
point at lattice distance one added on either side of $\SDelta_3$.
There can be different models of this type, depending on where the line
joining the two extra points intersects $\SDelta_3$. The largest $\Delta$
always occurs when this intersection point is the tip of $\SDelta_3$.
We find the following types:
\vspace*{-6pt}
\begin{itemize}\itemsep1pt
\item $E_8\times\{1\}$: there are 16 such cases belonging to the web 
studied in \sref{E81}, the largest one being of course $\cM_{11,491}$ 
at the upper left corner of the Hodge plot;

\item $E_7\times\{1\}$: this is the case of the web of \sref{E71} with dual 
models corresponding to $E_8{\times}SU(2)$; seven of these
polytopes are minimal with $\chi \ge -480$;

\item $E_8{\times}\{1'\}$: here $\SDelta_3$ consists of a 3-top of type $E_8$
together with a minimal bottom whose single point, however, is not the extension
of the long edge of the $E_8$ top; the dual slice $\Delta_3$ is
of the type $E_7'\times\{1\}$, where $E_7'$ is an alternative version of an 
$E_7$ type 3-top; there are 6 models of this type with $\chi \ge -480$, the 
largest such $\Delta$ corresponding to Hodge numbers $(12,318)$;

\item $F_4{\times}\{1\}$: there are 3 models of this type (with $\Delta_3$ of
the type $E_8\times G_2$); the largest being the one considered above, with
Hodge numbers $(9,321)$;

\item $E_7{\times}\{1'\}$: there is one model of this type (analogous to 
$E_8\times\{1'\}$) with Hodge numbers $(11,251)$; 
here the polytope $\Delta_3$ allows two interpretations: $E_7'{\times}SU(2)$
with a Weierstrass fiber as usual, or $E_8\times\{1\}$ with a fiber that is
a quartic curve in $\IP_{112}^2$;

\item $E_6{\times}\{1\}$: the largest $\Delta$ of this type is not maximal,
being contained in the largest $\Delta$ of the $F_4\times\{1\}$ case,
but an alternative one (where the two points outside $\SDelta_3$ do not form 
an edge) gives rise to the Hodge numbers $(12,264)$;

\item $G_2{\times}\{1\}$: there is one model of this type whose dual corresponds 
to $E_8\times F_4$; the Hodge numbers are $(7,271)$;
\item $SO(10){\times}\{1\}$: there is one model with Hodge numbers $(8,250)$;
$\Delta_3$ can be sliced in two distinct ways: one is the expected 
$E_8{\times}SU(4)$ and the other one is of the type $SO(24){\times}\{1\}$;

\item $\{1\}{\times}\{1\}$: there are 2 models, both with Hodge numbers 
$(3,243)$, corresponding to Weierstrass fibrations over $\IP_{112}^2$ or over
$\IP^1\times\IP^1$, respectively; in both cases $\Delta_3$ is the polytope
that can be interpreted as $E_8\times E_8$ or $SO(32){\times}\{1\}$.
\end{itemize}

Finally there are two models that do not exhibit $K3$ fibration structures:
\vspace*{-6pt}
\begin{itemize}\itemsep1pt
\item a model with Hodge numbers $(5,251)$ that admits a $\IZ_2$ quotient,
the resulting polytope corresponding to $F_4{\times}G_2$;

\item the Weierstrass elliptic fiber over the base $\IP^2$, with Hodge numbers $(2,272)$.
\end{itemize}

Let us also note that the largest $\Delta$ (with 311 points) that does not 
correspond to a Weierstrass fibration nevertheless corresponds to a dual pair
of nested fibrations, with $\SDelta_3$ of the type 
$E_7{\times}\{1\}$ with fiber in $\IP_{112}^2$ and $\Delta_3$ of the type 
$E_7{\times}\{1\}$ with fiber in the dual of $\IP_{112}^2$.
\newpage
\section*{Acknowledgements}
This project arose from the discussions which followed a talk given by PC at the Kreuzer Memorial Conference. PC wishes to thank the Conference organizers and the Schr\"odinger Institute and also the Perimeter Institute and ICTP, Trieste, for support and hospitality while he was engaged on this work. AC wishes to thank the University College, Oxford and STFC for supporting his graduate studies. We would like to thank also Helge Ruddat for his early participation in the project and for helpful discussions. 
\vskip30pt
\bibliography{bibfile}{}

\begin{thebibliography}{10}

\bibitem{Batyrev:1993dm}
Victor~V. Batyrev.
\newblock {Dual Polyhedra and Mirror Symmetry for Calabi-Yau Hypersurfaces in
  Toric Varieties}.
\newblock {\em J.Alg.Geom.}, 3, 1994.

\bibitem{Kreuzer:2000xy}
Maximilian Kreuzer and Harald Skarke.
\newblock {Complete classification of reflexive polyhedra in four-dimensions}.
\newblock {\em Adv.Theor.Math.Phys.}, 4:1209--1230, 2002.

\bibitem{Kreuzer:2000qv}
Maximilian Kreuzer and Harald Skarke.
\newblock {Reflexive polyhedra, weights and toric Calabi-Yau fibrations}.
\newblock {\em Rev.Math.Phys.}, 14:343--374, 2002.

\bibitem{Avram:1996pj}
A.C. Avram, M.~Kreuzer, M.~Mandelberg, and H.~Skarke.
\newblock {Searching for K3 fibrations}.
\newblock {\em Nucl.Phys.}, B494:567--589, 1997.

\bibitem{Candelas:1996su}
Philip Candelas and Anamaria Font.
\newblock {Duality between the webs of heterotic and type II vacua}.
\newblock {\em Nucl.Phys.}, B511:295--325, 1998.

\bibitem{Candelas:1997pq}
Philip Candelas and Harald Skarke.
\newblock {F theory, SO(32) and toric geometry}.
\newblock {\em Phys.Lett.}, B413:63--69, 1997.

\bibitem{Candelas:1997eh}
Philip Candelas, Eugene Perevalov, and Govindan Rajesh.
\newblock {Toric geometry and enhanced gauge symmetry of F theory / heterotic
  vacua}.
\newblock {\em Nucl.Phys.}, B507:445--474, 1997.

\bibitem{Bouchard:2003bu}
Vincent Bouchard and Harald Skarke.
\newblock {Affine Kac-Moody algebras, CHL strings and the classification of
  tops}.
\newblock {\em Adv.Theor.Math.Phys.}, 7:205--232, 2003.

\bibitem{0813.14039}
William Fulton.
\newblock {\em {Introduction to toric varieties. The 1989 William H. Roever
  lectures in geometry.}}
\newblock {Annals of Mathematics Studies. 131. Princeton, NJ: Princeton
  University Press. xi, 157 p.}, 1993.

\bibitem{1223.14001}
David~A. Cox, John~B. Little, and Henry~K. Schenck.
\newblock {\em {Toric varieties.}}
\newblock {Graduate Studies in Mathematics 124. Providence, RI: American
  Mathematical Society (AMS). xxiv, 841~p. }, 2011.

\bibitem{Skarke:1998yk}
Harald Skarke.
\newblock {String dualities and toric geometry: An Introduction}.
\newblock {\em Chaos Solitons Fractals}, 1998.

\bibitem{Klemm:1995tj}
A.~Klemm, W.~Lerche, and P.~Mayr.
\newblock {K3 Fibrations and heterotic type II string duality}.
\newblock {\em Phys.Lett.}, B357:313--322, 1995.

\bibitem{Vafa:1995gm}
Cumrun Vafa and Edward Witten.
\newblock {Dual string pairs with N=1 and N=2 supersymmetry in
  four-dimensions}.
\newblock {\em Nucl.Phys.Proc.Suppl.}, 46:225--247, 1996.

\bibitem{Aspinwall:1995vk}
Paul~S. Aspinwall and Jan Louis.
\newblock {On the ubiquity of K3 fibrations in string duality}.
\newblock {\em Phys.Lett.}, B369:233--242, 1996.

\bibitem{Hosono:1996ua}
S.~Hosono, B.H. Lian, and Shing-Tung Yau.
\newblock {Calabi-Yau varieties and pencils of K3 surfaces}.
\newblock 1996.

\bibitem{Perevalov:1997vw}
Eugene Perevalov and Harald Skarke.
\newblock {Enhanced gauged symmetry in type II and F theory compactifications:
  Dynkin diagrams from polyhedra}.
\newblock {\em Nucl.Phys.}, B505:679--700, 1997.

\bibitem{Kreuzer:1997zg}
Maximilian Kreuzer and Harald Skarke.
\newblock {Calabi-Yau four folds and toric fibrations}.
\newblock {\em J.Geom.Phys.}, 26:272--290, 1998.

\bibitem{Taylor:2012dr}
Washington Taylor.
\newblock {On the Hodge structure of elliptically fibered Calabi-Yau
  threefolds}.
\newblock 2012.

\end{thebibliography}
\bibliographystyle{unsrt}
\end{document}